\documentclass[3p,times]{elsarticle}

\usepackage{amssymb}
\usepackage{amsmath}

\usepackage[figuresright]{rotating}

\begin{document}

\begin{frontmatter}




\title{A fully implicit, conservative, non-linear, electromagnetic hybrid particle-ion/fluid-electron algorithm}


\author{A.~Stanier, L. Chac\'{o}n, G. Chen}

\address{Applied Mathematics and Plasma Physics, Los Alamos National Laboratory, Los Alamos, NM 87545, United States}

\begin{abstract}

The quasi-neutral hybrid model with kinetic ions and fluid electrons is a promising approach for bridging the inherent multi-scale nature of many problems in space and laboratory plasmas. Here, a novel, implicit, particle-in-cell based scheme for the hybrid model is derived for multi-dimensional electromagnetic problems with multiple ion species, which features global mass, momentum and energy conservation. The scheme includes sub-cycling and orbit averaging of the ions, with cell-centered finite differences and implicit midpoint time advance. To reduce discrete particle noise, the scheme allows arbitrary-order shape functions for the particle-mesh interpolations and the application of conservative binomial smoothing. The algorithm is verified for a number of test problems to demonstrate the correctness of the implementation, the unique conservation properties, and the favorable stability properties of the new scheme. In particular, there is no indication of unstable growth of the finite-grid instability for a population of cold ions drifting through a uniform spatial mesh, in a set-up where several commonly used non-conservative schemes are highly unstable.

\end{abstract}




\end{frontmatter}


\section{Introduction}
\label{}


Hybrid plasma models, in which some component of the plasma is treated kinetically and a fluid description is used for the remainder, are promising candidates to model multi-scale plasma phenomena~\cite{winske03,lipatov02,karimabadi14, stanier15prl, sturdevant17}. Hybrid models typically feature higher fidelity than reduced fluid descriptions, while remaining more computationally feasible than solving full kinetic models. A variety of hybrid models are discussed in the literature, which differ mainly by the choice of the kinetic component and the calculation of the electric field. Here, the term will apply to the quasi-neutral model with full-orbit kinetic ions and fluid electrons. Such a model is valid for non-relativistic plasmas where the light wave is eliminated, and the electric field is calculated with Ohm's law rather than solving the full Maxwell equations. The kinetic equation for the ions is solved using the Particle-In-Cell (PIC) approach~\cite{birdsall91, hockney81} where a large number of discrete macro-particles trace out the evolution of the distribution function using the method of characteristics. This approach avoids the need to solve for the distribution function on a 6D (3D-3V) grid and can be highly optimized to run on modern computer architectures with threading and vectorization, but suffers from discrete particle noise that scales as $\sqrt{N}$ with the finite number of macro-particles $N$.

There is significant literature on the development of numerical algorithms based on the hybrid PIC ion and fluid electron model, driven by the applications of space weather modeling and magnetic fusion research, see Refs.~\cite{winske03,lipatov02} for reviews. The majority of existing codes employ explicit time-stepping, and much of the development has focused on key algorithmic issues related to accuracy and stability of the time integration schemes. One issue arises from the use of leapfrog time advance where the particle positions and momenta are staggered in time, while statically evaluating Ohm's law for the electric field using these as sources. This results in an interdependence in which the particle velocity update and the calculation of the electric field depend implicitly on each other~\cite{winske03, karimabadi04}. To make this explicit, some form of the predictor-corrector method~\cite{harned82, winske86, kunz14}, moment method~\cite{winske88,matthews94}, or extrapolation~\cite{fujimoto90,thomas90,karimabadi04} has been used previously. A further issue with explicit methods concerns the stable propagation of the dispersive whistler waves that are present in the hybrid model. In the absence of dissipation, it has been shown that two-step explicit schemes such as the second-order Runge-Kutta method are unconditionally unstable to whistler waves~\cite{kunz13,kunz14}. For stability in higher-order explicit methods, it is necessary to resolve the quadratic CFL-limited time-step associated with dispersive Whistler waves, $\Delta t \propto n (\Delta x)^2/B$, where $\Delta t$ is the time-step, $\Delta x$ the grid spacing, $n$ the density, and $B$ the magnetic field strength. In practice, the field solve is often sub-cycled to satisfy this condition while avoiding the expense of a particle push every time-step~\cite{swift95,matthews94,karimabadi04}. This condition can be particularly prohibitive in near-vacuum regions, and several schemes have been proposed to avoid associated numerical instabilities~\cite{hewett80,harned82,amano14}.

Implicit time-stepping has also been explored within hybrid algorithms. Early work focused on implicit time-stepping only for the field solve~\cite{hewett80, lipatov02}, but recent papers have also used an implicit particle advance~\cite{cheng13,sturdevant16b,sturdevant17}. In particular, Ref.~\cite{sturdevant16b} applied the techniques of ion sub-cycling and orbit averaging to an implicit electrostatic $\delta f$ hybrid model. This was then extended to toroidal geometry in Ref.~\cite{sturdevant17}, where it was successfully used to model the low-frequency ion-temperature-gradient instability with full-orbit kinetic ions. One of the key successes of implicit particle methods for the full kinetic PIC model is the capability for discrete energy conservation using a finite time-step~\cite{chen11,markidis11} (the earlier explicit method of Ref.~\cite{lewis70} conserved energy only in the limit of $\Delta t\rightarrow 0$). It has been demonstrated that discrete energy conservation can be found in addition to local charge conservation in multi-dimensional~\cite{chen15} curvilinear geometries~\cite{chacon16} with sub-cycling and orbit averaging. Recently, there has been particular interest in discrete methods that preserve geometric structures of the Vlasov-Maxwell system~\cite{squire12,evstatiev13,crouseilles15,qin16,kraus17}.

An example of the macroscopic consequences of violating discrete conservation in fully kinetic PIC algorithms is the finite-grid instability, caused by aliasing errors between the particles, which live in continuous space, and the discrete spatial grid. A dramatic example of this is the numerical instability of a cold beam moving through a charge neutralizing background~\cite{birdsall91}, which would be stable in the real physical system. The instability is manifest even in numerical schemes that discretely conserve either momentum or energy, where it results in a large violation of the quantity that is not discretely conserved~\cite{birdsall91,okuda72,markidis11}. In principle, simultaneous momentum and energy conservation could suppress such an instability, but no such scheme that uses a spatial grid presently exists~\cite{brackbill16}. Grid-less particle methods, which use a Fourier basis for the fields instead of a spatial grid, are able to conserve both~\cite{vlad01,evstatiev13} and are stable~\cite{huang16}. For the quasi-neutral hybrid model, an analogous finite grid instability has been demonstrated to occur for a momentum-conserving scheme~\cite{rambo95}. In this case the threshold for instability occurs for a cold but finite ion-to-electron temperature ratio $0<T_i/T_e < 1$, where the exact value depends on the order of the interpolation scheme used.  

In this paper, we derive a novel discrete hybrid particle-kinetic-ion and fluid-electron model that features global mass, momentum, and energy conservation simultaneously. The fully implicit, non-linear, and electromagnetic model is presented in a multi-dimensional form for multiple species of ions and a mass-less electron fluid with scalar pressure. Different implementations of the model using Jacobian-free Newton-Krylov methods are discussed, which differ by the non-linear convergence of the momentum error and by algorithmic efficiency. The implementation that is suitable for multi-scale simulations allows sub-cycling for accurate integration of the ion orbits, and collecting orbit-averaged moments to reduce discrete particle noise. The favorable stability properties of the algorithm are demonstrated for the problem of a cold ion beam moving through the spatial mesh~\cite{rambo95}. For a particularly challenging set-up ($T_i/T_e = 1/600$ with nearest-grid-point interpolation and without smoothing), there is no indication of the finite-grid instability that is present in several commonly used non-energy-conserving schemes. Further verification tests are then presented, ranging from 1D-1V electrostatic problems to 2D-3V non-linear electromagnetic tests, to demonstrate the correctness of the implementation.

\section{Hybrid kinetic-ion fluid-electron plasma model}

There are a number of options for the choice of the electron fluid model, and it is appropriate to comment firstly on the choices that have been made here. Finite electron inertia has been included in several hybrid algorithms~\cite{winske03,cheng13,munoz16}, where it can mitigate the stiffness of the whistler wave at short wavelength and in near vacuum regions. In such algorithms, electron inertia is often implemented with a reduced value of the ion-to-electron mass ratio to reduce the separation between the ion and electron scales. The electron pressure $p_e$ in hybrid models is commonly taken to be either isothermal, $p_e=T_{e0} n$, or to use an adiabatic equation of state, $p_e = T_{e0} n^{\gamma}$, where $T_{e0}$ is the initial temperature and $\gamma$ is the ratio of specific heats. For modeling magnetic reconnection, notable extensions include the implementation of anisotropic electron pressure~\cite{le16,ohia12} and even off-diagonal pressure tensor effects~\cite{winske03} using an \textit{ad-hoc} closure. Here, to demonstrate the core algorithm, we choose a simple electron model with mass-less fluid electrons and a scalar electron pressure, and leave extensions to future work. One of the intended applications of our algorithm is the modeling of magnetic reconnection, in which ion kinetic effects can be of primary importance~\cite{karimabadi04,stanier15prl,ng15}. For reconnection, it is necessary to include some term to balance the reconnection electric field at the X-point. Here, we include the options to use plasma resistivity to break the electron frozen-in condition, and also electron viscosity that can provide dissipation in a narrow band in k-space to target only the smallest scales. For energy conservation in the presence of dissipative terms in Ohm's law, however, it is necessary to use a separate electron pressure evolution equation with corresponding heating terms.

The continuum hybrid description that is used here is given by the following equations

\begin{equation}\label{vlasov}\partial_t f_s + \boldsymbol{\nabla} \cdot \left(f_s\boldsymbol{v}\right) + (q_s/m_s)\left(\boldsymbol{E}^* + \boldsymbol{v} \times \boldsymbol{B}\right) \cdot \boldsymbol{\nabla}_v f_s = 0,\end{equation}
\begin{equation}\label{faraday}\partial_t \boldsymbol{B} = - \boldsymbol{\nabla} \times \boldsymbol{E}, \end{equation}
\begin{equation}\label{ohms}\boldsymbol{E} = \boldsymbol{E}^* + \eta \boldsymbol{j} = -\boldsymbol{u}_{i} \times \boldsymbol{B} + \frac{\boldsymbol{j}\times \boldsymbol{B}}{ne} - \frac{\boldsymbol{\nabla}p_e}{ne} -\frac{\boldsymbol{\nabla} \cdot \overleftrightarrow{\Pi}_e}{ne} + \eta \boldsymbol{j},\end{equation}
\begin{equation}\label{epress}\left(\gamma -1\right)^{-1} \left[\partial_t p_e + \boldsymbol{\nabla} \cdot \left(\boldsymbol{u}_e p_e\right)\right] + p_e \boldsymbol{\nabla}\cdot \boldsymbol{u}_e = H_e - \boldsymbol{\nabla} \cdot \boldsymbol{q}_e,\end{equation}
where $f_s$ is the distribution function for ion species s, $\boldsymbol{v}$ is the velocity space co-ordinate, $\boldsymbol{E}^*$ is the electric field ($\boldsymbol{E}$) minus the collisional friction (see below), $\boldsymbol{B}$ is the magnetic field, $n=n_e=\frac{1}{e}\sum_s q_s\int f_s d^3v=\frac{1}{e}\sum_s q_s n_s$ is the quasi-neutral number density, $\boldsymbol{u}_i=\left(\sum_s q_s \int f_s \boldsymbol{v}d^3v\right)/en= \left(\sum_s q_s n_s\boldsymbol{u}_s\right)/en$ is the ion current carrying drift velocity,  $p_e$ is the electron pressure, $\boldsymbol{j} = \boldsymbol{\nabla}\times \boldsymbol{B}/\mu_0$ is the current density, $\boldsymbol{u}_e = \boldsymbol{u}_i - \boldsymbol{j}/ne$ is the bulk electron velocity, $\overleftrightarrow{\Pi}_e = -\mu_e \boldsymbol{\nabla}\boldsymbol{u}_e$ is the electron viscous tensor, $\boldsymbol{q}_e = -\kappa_e \boldsymbol{\nabla}(p_e/n)$ is the electron heat flux, and $H_e=\eta j^2 - \overleftrightarrow{\Pi}_e:\boldsymbol{\nabla}\boldsymbol{u}_e$ is the Ohmic and electron viscous heating. The coefficients are the charge and mass of species $s$, $q_s$ and $m_s$ respectively, the ratio of specific heats $\gamma$, the resistivity $\eta$, the electron viscosity $\mu_e$, and the electron heat conductivity $\kappa_e$.

\subsection{Conservation properties in the continuum}

\subsubsection{Mass conservation and quasi-neutrality}

Taking the zeroth moment of the Vlasov equation~(\ref{vlasov}) gives the local mass conservation equation for ions of species $s$
\begin{equation}\label{continuummass}\partial_t n_s + \boldsymbol{\nabla}\cdot \left( n_s\boldsymbol{u}_s \right) = 0.\end{equation}
Due to quasi-neutrality there is no need to solve a separate electron fluid mass conservation equation. Instead, this property requires the ambipolarity condition to be satisfied for consistency. Multiplying Eq.~(\ref{continuummass}) by the ion charge $q_s$, taking the sum over species $s$, and using the definitions of $n$, $\boldsymbol{u}_i$, $\boldsymbol{u}_e$, and $\boldsymbol{j}$, gives 
\begin{equation}\label{consistency}\boldsymbol{\nabla}\cdot \left(e n \left(\boldsymbol{u}_i - \boldsymbol{u}_e\right) \right) =  \boldsymbol{\nabla} \cdot \boldsymbol{j} = \boldsymbol{\nabla}\cdot \boldsymbol{\nabla}\times \boldsymbol{B}/\mu_0 = 0.\end{equation}

\subsubsection{Momentum conservation}

The ion momentum density equation for species $s$ is found from the first moment of the Vlasov equation~(\ref{vlasov}). 
\begin{equation}\label{continuummomentum}\partial_t \left(m_s n_s \boldsymbol{u}_s\right) + \boldsymbol{\nabla}\cdot \overleftrightarrow{\mathbb{P}}_s = q_s n_s \left(\boldsymbol{E}^* + \boldsymbol{u}_s \times \boldsymbol{B}\right), \end{equation}
where $\overleftrightarrow{\mathbb{P}}_s = \int m_s \boldsymbol{v}\boldsymbol{v} f_s d^3v$ is the pressure tensor for species $s$. Taking the sum over species $s$ gives the total ion momentum
\begin{align*}\partial_t \left(\sum_s m_s n_s \boldsymbol{u}_s\right) + \boldsymbol{\nabla} \cdot \left(\sum_s  \overleftrightarrow{\mathbb{P}}_s\right) &= \left(\sum_s q_sn_s\right)\boldsymbol{E}^* + \left(\sum_s q_s n_s \boldsymbol{u}_s\right) \times \boldsymbol{B}\\
&=en\left(\boldsymbol{E}^* + \boldsymbol{u}_i \times \boldsymbol{B}\right).\end{align*}
Using the definition of $\boldsymbol{E}^*$ in Eq.~(\ref{ohms}), and the identity $(\boldsymbol{\nabla}\times \boldsymbol{B}) \times \boldsymbol{B} = \boldsymbol{\nabla} \cdot [\boldsymbol{B}\boldsymbol{B} - (B^2/2) \mathbb{I}]$ gives the local momentum density conservation equation
\begin{equation}\partial_t \left(\sum_s m_s n_s \boldsymbol{u}_s\right) + \boldsymbol{\nabla} \cdot \left[\sum_s  \overleftrightarrow{\mathbb{P}}_s + \overleftrightarrow{\Pi}_e +  \left(p_e + \frac{B^2}{2\mu_0}\right)\mathbb{I} - \frac{\boldsymbol{B}\boldsymbol{B}}{\mu_0}\right] = \boldsymbol{0}.\end{equation} 
Here, the reason for using the frictionless electric field $\boldsymbol{E}^*$ becomes clear: using the total electric field would give a frictional source of total momentum. The reason for this is the \textit{ad-hoc} but widely used~\cite{swift95,lipatov02,winske03,kunz13,cheng13} choice to include resistive friction (due to ion-electron collisions) in Ohm's law~(\ref{ohms}) without an explicit collision operator in the kinetic ion equation~(\ref{vlasov}). We follow this choice here, and leave consideration of self-consistent ion-electron collisions for future work. 

The total momentum $\int \sum_s m_s n_s \boldsymbol{u}_s dV$ is constant in a domain $V$ only for suitable boundary conditions, where the flux terms vanish. Examples include periodic boundaries, but exclude open boundaries with a net momentum flux, or conducting walls. 

\subsubsection{Energy conservation}

Taking the contracted second moment of the ion Vlasov equation gives an equation for $\kappa_s = \int m_s v^2 f_s d^3v/2$, the kinetic energy density for ions of species $s$
\begin{equation}\partial_t \kappa_s + \boldsymbol{\nabla} \cdot \boldsymbol{Q}_s = q_s n_s \boldsymbol{E}^* \cdot \boldsymbol{u}_s,\end{equation}
where $\boldsymbol{Q}_s = \int m_s v^2\boldsymbol{v}f_s d^3v/2$ is the energy flux. Taking the sum over species $s$, and using the definition of $\boldsymbol{u}_i$ gives
\begin{equation}\label{ionkinen}\partial_t \sum_s \kappa_s + \boldsymbol{\nabla} \cdot \left(\sum_s  \boldsymbol{Q}_s\right) = e n \boldsymbol{E}^* \cdot \boldsymbol{u}_i.\end{equation}

An equation for the magnetic energy density $B^2/2\mu_0$ (Poynting's theorem) can be found from the Faraday equation~(\ref{faraday})
\begin{equation}\label{magnetic}\partial_t \left(\frac{B^2}{2\mu_0}\right) = - \frac{\left(\boldsymbol{\nabla}\times \boldsymbol{E}\right) \cdot \boldsymbol{B}}{\mu_0} = - \boldsymbol{\nabla}\cdot \left(\frac{\boldsymbol{E}\times \boldsymbol{B}}{\mu_0}\right) - \boldsymbol{E}\cdot \boldsymbol{j},\end{equation}
which is the divergence of the Poynting flux and the magnetic dissipation respectively. Note the total electric field is used here. Taking the sum of the ion kinetic energy in Eq.~(\ref{ionkinen}), the magnetic energy in Eq.~(\ref{magnetic}) and the electron internal energy in Eq.~(\ref{epress}) gives the total energy

\begin{equation}\label{totalenergy}\partial_t \left[ \sum_s \kappa_s + \frac{p_e}{\gamma -1} + \frac{B^2}{2\mu_0}\right] + \boldsymbol{\nabla} \cdot \left[ \sum_s \boldsymbol{Q}_s + \frac{\gamma}{\gamma - 1} \boldsymbol{u}_e p_e + \overleftrightarrow{\Pi}_e\cdot \boldsymbol{u}_e + \boldsymbol{q}_e + \frac{\boldsymbol{E}\times \boldsymbol{B}}{\mu_0}\right] = 0.\end{equation}

Again, the total energy $\int [\sum_s \kappa_s + p_e/(\gamma-1) + B^2/2\mu_0] dV$ is conserved in a domain $V$ only for suitable boundary conditions, such as periodic or conducting walls (with zero viscous stress and heat-flux losses).

\section{A discrete, conservative formulation for the hybrid model}

\subsection{Discrete model}

Our discrete momentum-energy conserving scheme is constructed on the key attributes of a cell-centered discretization in space, and an implicit midpoint method in time. For this choice, special consideration must be taken in the discretization of advection operators to handle problems with strong flows. This will be discussed further below and in the numerical examples. Derivatives are approximated by central differences, which are defined as $(\delta_x \chi)_{ijk} = (\chi_{i+1jk} - \chi_{i-1jk})/(2\Delta x)$, the discrete gradient as $(\boldsymbol{\nabla}\chi)_{ijk} = (\delta_x \chi)_{ijk} \boldsymbol{\hat{x}} + (\delta_y \chi)_{ijk} \boldsymbol{\hat{y}} + (\delta_z \chi)_{ijk} \boldsymbol{\hat{z}}$, the discrete curl as $\left(\boldsymbol{\nabla}\times \boldsymbol{N}\right)_{ijk} = ((\delta_yN_z)_{ijk}-(\delta_zN_y)_{ijk})\boldsymbol{\hat{x}} +  ((\delta_zN_x)_{ijk}-(\delta_xN_z)_{ijk})\boldsymbol{\hat{y}}+ ((\delta_xN_y)_{ijk}-(\delta_yN_x)_{ijk})\boldsymbol{\hat{z}}$, and the discrete divergence as $(\boldsymbol{\nabla}\cdot \boldsymbol{N})_{ijk} = (\delta_xN_x)_{ijk} + (\delta_yN_y)_{ijk} + (\delta_zN_z)_{ijk}$. These discretizations naturally preserve many of the continuum properties of the discrete operators, such as $(\boldsymbol{\nabla}\cdot \left(\boldsymbol{\nabla}\times \boldsymbol{N}\right)_{ijk})_{ijk} = 0$, e.g.~\cite{chacon04}. For quantities $\chi$ that are known at integer time-step, the definition at the half-integer time-step is given as $\chi^{n+1/2} = \tfrac{1}{2}(\chi^{n+1} + \chi^n)$.


For the remainder of the paper, a vector potential formulation is adopted using the Weyl gauge ($\phi=0$), such that $\boldsymbol{B} = \boldsymbol{\nabla} \times \boldsymbol{A}$ and $\boldsymbol{E} = - \partial_t \boldsymbol{A}$. This choice is made for ease of implementation into an existing fluid framework, see below, but we note the conservation properties derived below and also the solenoidal condition $(\boldsymbol{\nabla} \cdot \boldsymbol{B})_{ijk} = 0$ hold for a magnetic field formulation discretized with the central differences defined above, e.g.~\cite{chacon04}. For clarity, we also leave the numerical description of the electron viscosity/hyper-resistivity to~\ref{hyperresistivity}.

The discrete Ohm's law and electron pressure equations are given by 
\begin{equation}\label{discreteA}-\left(\frac{\boldsymbol{A}_{ijk}^{n+1}-\boldsymbol{A}_{ijk}^{n}}{\Delta t}\right) = \boldsymbol{E}^{*,\,n+1/2}_{ijk} + \eta \boldsymbol{j}_{ijk}^{n+1/2},\end{equation}
\begin{equation}\label{discreteE}\boldsymbol{E}^{*,\,n+1/2}_{ijk} = - \boldsymbol{u}^{n+1/2}_{ijk} \times \left(\boldsymbol{\nabla}\times \boldsymbol{A}\right)_{ijk}^{n+1/2} + \frac{\boldsymbol{j}_{ijk}^{n+1/2}\times \left(\boldsymbol{\nabla}\times \boldsymbol{A}\right)^{n+1/2}_{ijk} - \left(\boldsymbol{\nabla} p_e\right)_{ijk}^{n+1/2}}{en_{ijk}^{n+1/2}},\end{equation}
\begin{equation}\label{discretePe}\frac{p_{e,ijk}^{n+1}-p_{e,ijk}^n}{\Delta t} + \gamma \left(\boldsymbol{\nabla} \cdot \left[\boldsymbol{u}_e p_e\right]\right)_{ijk}^{n+1/2} = \left(\gamma-1\right) \left[\boldsymbol{u}_{e,ijk}^{n+1/2} \cdot \left(\boldsymbol{\nabla}p_e\right)_{ijk}^{n+1/2} - \left(\boldsymbol{\nabla}\cdot \boldsymbol{q}_e\right)_{ijk}^{n+1/2} + \eta \left(j_{ijk}^{n+1/2}\right)^2\right].\end{equation}
Here, $\boldsymbol{j}_{ijk}^{n+1/2} = (\boldsymbol{\nabla}\times (\boldsymbol{\nabla} \times \boldsymbol{A}^{n+1/2}))_{ijk}/\mu_0$, the electron velocity is defined as $\boldsymbol{u}_{e,ijk}^{n+1/2} = \boldsymbol{u}_{ijk}^{n+1/2} - \boldsymbol{j}_{ijk}^{n+1/2}/en_{ijk}^{n+1/2}$, and $\boldsymbol{q}_{e,ijk}^{n+1/2} = -\kappa_e \left[\boldsymbol{\nabla}(p_{e,ijk}^{n+1/2}/n_{ijk}^{n+1/2})\right]_{ijk}$. The definitions of $n_{ijk}^{n+1/2}$ and $\boldsymbol{u}_{ijk}^{n+1/2}$ are given below.

For clarity, the conservation properties will be firstly demonstrated using the same time-step for the particle equations of motion as for the electromagnetic field and electron pressure equations. This is then generalized to include particle sub-cycling with orbit-averaged moments below. Also, from hereon, the species index $s$ is incorporated into the macro-particle index $p$ without loss of generality. The equations of motion for the macro-particle $p$ with charge $q_p$ and mass $m_p$ are given by
\begin{equation}\label{discreteX}\frac{\boldsymbol{x}_{p}^{n+1} - \boldsymbol{x}_{p}^n}{\Delta t} = \boldsymbol{v}_{p}^{n+1/2},\end{equation}
\begin{equation}\label{discreteV}\frac{\boldsymbol{v}_{p}^{n+1} - \boldsymbol{v}_{p}^n}{\Delta t} = \frac{q_p}{m_p} \left(\boldsymbol{E}_{p}^{*,\,n+1/2} + \boldsymbol{v}_{p}^{n+1/2} \times \boldsymbol{B}_{p}^{n+1/2}\right),\end{equation}
where $\boldsymbol{x}_{p}^n, \boldsymbol{v}_{p}^n$ is the known position and velocity of particle $p$ at time $t=n\Delta t$, and $\boldsymbol{B}_{p}^{n+1/2}$, $\boldsymbol{E}_{p}^{*,\,n+1/2}$ are the magnetic and frictionless electric fields interpolated to the particle position at $t=(n+1/2)\Delta t$ as 
\begin{equation}\label{scatterB}\boldsymbol{B}_{p}^{n+1/2} = \sum_{ijk} S\left(x_i - x_{p}^{n+1/2}\right) S\left(y_j - y_{p}^{n+1/2}\right) S\left(z_k - z_{p}^{n+1/2}\right)\left(\boldsymbol{\nabla}\times \boldsymbol{A}\right)_{ijk}^{n+1/2},\end{equation}
\begin{equation}\label{scatterE}\boldsymbol{E}_{p}^{*,\,n+1/2} = \sum_{ijk} S\left(x_i - x_{p}^{n+1/2}\right) S\left(y_j - y_{p}^{n+1/2}\right) S\left(z_k - z_{p}^{n+1/2}\right) \boldsymbol{E}^{*,\,n+1/2}_{ijk},\end{equation}
where $S(x_i-x_p)$ is, for now, an arbitrary order shape function for the particle-grid interpolations~\cite{birdsall91}.

Finally, the moments $n_{ijk}^{n+1/2}$ and $\boldsymbol{u}^{n+1/2}_{ijk}$ are calculated from the macro-particles and used in Eqs.~(\ref{discreteE})-(\ref{discretePe}) to close the set of equations, as
\begin{equation}\label{nmoment}n_{ijk}^{n+1/2} = \frac{1}{\Delta V}\sum_p \frac{q_p}{e} S\left(x_i - x_{p}^{n+1/2}\right) S\left(y_j - y_{p}^{n+1/2}\right) S\left(z_k - z_{p}^{n+1/2}\right),\end{equation}
\begin{equation}\label{umoment}\boldsymbol{u}^{n+1/2}_{ijk} =\frac{1}{n_{ijk}^{n+1/2} \,\Delta V}\sum_p \frac{q_p}{e} S\left(x_i - x_{p}^{n+1/2}\right) S\left(y_j - y_{p}^{n+1/2}\right) S\left(z_k - z_{p}^{n+1/2}\right) \boldsymbol{v}_{p}^{n+1/2},\end{equation}
where $\Delta V = \Delta x \Delta y \Delta z$ is the cell volume.

\subsection{Discrete mass and quasi-neutrality}

As is typical for particle methods, the particle mass, $m_p$, and charge, $q_p$, are constant along the characteristic particle trajectories, and are thus locally conserved in the Lagrangian sense. To conserve mass and charge globally it is then sufficient to conserve the total number of particles, which can be done for suitable boundary conditions in the absence of particle sources and sinks. 

In the fully kinetic Vlasov-Maxwell system it is important to conserve charge locally on the spatial mesh, because small truncation errors in the local charge conservation equation can accumulate to give significant violations in Gauss' law (see Refs.~\cite{villasenor92,mardahl97} for discussions). For the quasi-neutral hybrid model, Gauss' law has been removed from the continuum equations, and instead the discrete form of Eq.~(\ref{consistency}) needs to be satisfied locally at the locations where the density moment is collected. This is satisfied for our discrete model, from the definition of $\boldsymbol{u}_{e,ijk}^{n+1/2}$ and the discrete divergence and curl operators that give $(\boldsymbol{\nabla}\cdot \boldsymbol{j})_{ijk}^{n+1/2} = 0$.

It is possible to gather density and momentum such that they satisfy a local mass conservation equation on the spatial grid using techniques developed for the full Vlasov-Maxwell model, e.g. Refs.~\cite{villasenor92,esirkepov01,chen11,chen15}. \ref{massconsv} describes a generalization of the method of Refs.~\cite{chen11,chen15} for three spatial dimensions, where the density and momentum measures are defined at different locations on a staggered grid. We have chosen not to directly use such measures here, as we find they do not give momentum conservation for the hybrid model. However, as described in Section~\ref{subcycleorbitav}, we use the same particle sub-stepping scheme as Refs.~\cite{chen11,chen15} where the ions strictly deposit contributions to the density and momentum at all of the cells they pass through in a given time-step. The cell-centered measures we use lie within truncation error from the locally-mass-conserving measures, and can not increase unboundedly due to local mass conservation in the Lagrangian sense. Since global mass conservation also holds, these truncation errors cancel when integrated over the whole domain. Although we typically use quadratic spline shape functions~(\ref{s2shape1},\ref{s2shape2}), the derivations below hold for arbitrary order shape functions and, as such, we drop the subscript.

\subsection{Discrete momentum conservation}

We define the total momentum for all particles (of all species) at time $t=n\Delta t$ by $\boldsymbol{P}^n = \sum_p m_p \boldsymbol{v}_{p}^n$. The discrete rate of change in total momentum is then given by \begin{align}\label{momderiv}\frac{\boldsymbol{P}^{n+1} - \boldsymbol{P}^n}{\Delta t} &=\sum_p m_p \frac{\boldsymbol{v}_{p}^{n+1} - \boldsymbol{v}_{p}^n}{\Delta t} = \sum_p q_p \left(\boldsymbol{E}^{*,\,n+1/2}_{p} + \boldsymbol{v}_{p}^{n+1/2}\times \boldsymbol{B}_{p}^{n+1/2}\right)\\
&=\sum_p q_p \sum_{ijk}S(x_i-x_{p}^{n+1/2})S(y_j-y_{p}^{n+1/2})S(z_k-z_{p}^{n+1/2})\left[\boldsymbol{E}^{*,\,n+1/2}_{ijk} + \boldsymbol{v}_{p}^{n+1/2}\times \left(\boldsymbol{\nabla}\times \boldsymbol{A}^{n+1/2}\right)_{ijk}\right] \nonumber \\
&=\sum_{ijk}  \Delta V \left[en_{ijk}^{n+1/2}\boldsymbol{E}^{*,\,n+1/2}_{ijk} +  en_{ijk}^{n+1/2}\boldsymbol{u}^{n+1/2}_{ijk}\times \left(\boldsymbol{\nabla}\times \boldsymbol{A}^{n+1/2}\right)_{ijk}\right] \nonumber \\ 
&=\sum_{ijk} \Delta V \left[\boldsymbol{j}_{ijk}^{n+1/2}\times \left(\boldsymbol{\nabla}\times \boldsymbol{A}\right)^{n+1/2}_{ijk} - \left(\boldsymbol{\nabla} p_e\right)_{ijk}^{n+1/2}\right], \nonumber
\end{align}
using the velocity update equation~(\ref{discreteV}), the definitions of $\boldsymbol{B}^{n+1/2}_{p}$ and $\boldsymbol{E}^{*,\,n+1/2}_{p}$ in Eqs.~(\ref{scatterB})-(\ref{scatterE}), collecting the moments $n_{ijk}^{n+1/2}$ and $\boldsymbol{u}_{ijk}^{n+1/2}$ as defined in Eqs.~(\ref{nmoment})-(\ref{umoment}), and using Ohm's law from Eq.~(\ref{discreteE}).
For suitable boundaries, e.g. periodic, we can use the discrete integration by parts (telescoping the sums). That is, for discrete scalar fields $\chi_{ijk}$, $\xi_{ijk}$, $\sum_{ijk} (\delta_x \chi)_{ijk} \xi_{ijk} = - \sum_{ijk} \chi_{ijk} (\delta_x \xi)_{ijk}$. Using this property, and noting that the sum over gradients vanishes in periodic domains, gives
\begin{align*}\frac{\boldsymbol{P}^{n+1}-\boldsymbol{P}^n}{\Delta t} = \sum_{ijk} \Delta V \left\{ - \left(\boldsymbol{\nabla}\times \boldsymbol{A}^{n+1/2}\right)_{ijk}\left[\boldsymbol{\nabla} \cdot \left(\boldsymbol{\nabla} \times \boldsymbol{A}^{n+1/2}\right)_{ijk}\right]_{ijk}\right\} = \boldsymbol{0},\end{align*}
from the definitions of the discrete divergence and curl operators given above.

In the above, it has been shown that the total momentum of the system is conserved in the sense that it has no contribution from spatial or temporal truncation error. In practice, there are two other sources of numerical error in an implicit algorithm: numerical round-off error due to finite precision arithmetic, and non-linear convergence error associated with an iterative implicit scheme. It will be shown below that the largest of these errors depends on the precise implementation within the non-linear solver.



\subsection{Discrete energy conservation}

The total ion kinetic energy at $t=n\Delta t$ is defined as $K_i^n = \sum_p m_p (v_{p}^{n})^2/2$. Then the discrete rate of change of $K_i$ is given by
\begin{align*}\frac{K_{i}^{n+1} - K_{i}^n}{\Delta t} &= \sum_p m_p \frac{\left(v_{p}^{n+1}\right)^2 - \left(v_{p}^{n}\right)^2}{2 \Delta t} \\
&= \sum_p m_p \boldsymbol{v}_{p}^{n+1/2} \cdot \left(\frac{\boldsymbol{v}_{p}^{n+1} - \boldsymbol{v}_{p}^{n}}{\Delta t}\right)\\
&= \sum_p q_p \boldsymbol{v}_{p}^{n+1/2} \cdot \boldsymbol{E}^{*,\,n+1/2}_{p}\\
&= \sum_p q_p \sum_{ijk} \boldsymbol{v}_{p}^{n+1/2} \cdot \boldsymbol{E}^{*,\,n+1/2}_{ijk}S\left(x_i - x_{p}^{n+1/2}\right) S\left(y_j - y_{p}^{n+1/2}\right) S\left(z_k - z_{p}^{n+1/2}\right)\\
&=\sum_{ijk}\Delta V  \left(en_{ijk}^{n+1/2}\boldsymbol{u}^{n+1/2}_{ijk}\right)\cdot \boldsymbol{E}^{*,\,n+1/2}_{ijk}\\
&=\sum_{ijk}\Delta V\,  \boldsymbol{u}^{n+1/2}_{ijk}\cdot \left[\boldsymbol{j}_{ijk}^{n+1/2}\times \left(\boldsymbol{\nabla}\times \boldsymbol{A}\right)^{n+1/2}_{ijk} - \left(\boldsymbol{\nabla} p_e\right)_{ijk}^{n+1/2}\right],
\end{align*}
using Eqs.~(\ref{discreteV}),~(\ref{scatterE}),~(\ref{umoment}), and Ohm's law from Eq.~(\ref{discreteE}).

We define the sum of magnetic energy at cell centers and integer time-step as $W_B^n = \sum_{ijk}\Delta V \left[(\boldsymbol{\nabla}\times \boldsymbol{A}^{n})_{ijk}\right]^2/2\mu_0$. The discrete rate of change of total magnetic energy is then

\begin{align*}\frac{W_B^{n+1}-W_B^n}{\Delta t} &= \sum_{ijk} \Delta V \frac{\left[(\boldsymbol{\nabla}\times \boldsymbol{A}^{n+1})_{ijk}\right]^2-\left[(\boldsymbol{\nabla}\times \boldsymbol{A}^{n})_{ijk}\right]^2}{2 \mu_0 \Delta t } \\
&=\sum_{ijk} \Delta V \left(\boldsymbol{\nabla}\times \boldsymbol{A}^{n+1/2}\right)_{ijk} \cdot \boldsymbol{\nabla} \times \left(\frac{\boldsymbol{A}^{n+1}_{ijk} - \boldsymbol{A}^n_{ijk}}{\mu_0 \Delta t}\right)\\
&=\sum_{ijk} \Delta V \left[\boldsymbol{j}_{ijk}^{n+1/2} \cdot \left(\frac{\boldsymbol{A}^{n+1}_{ijk} - \boldsymbol{A}^n_{ijk}}{\Delta t}\right)\right]\\
&=\sum_{ijk} \Delta V\left[\boldsymbol{j}_{ijk}^{n+1/2} \cdot \boldsymbol{u}^{n+1/2}_{ijk} \times \left(\boldsymbol{\nabla}\times \boldsymbol{A}\right)_{ijk}^{n+1/2}+ \frac{\boldsymbol{j}_{ijk}^{n+1/2} \cdot  \left(\boldsymbol{\nabla} p_e\right)_{ijk}^{n+1/2}}{en_{ijk}^{n+1/2}} - \eta \left(j_{ijk}^{n+1/2}\right)^2\right],\end{align*}
by discrete integration by parts, summing fluxes to zero, and using Eq.~(\ref{discreteA}).

Since we neglect electron inertia, the sum of electron kinetic energy (electron thermal energy) is defined as $K_e^n = \sum_{ijk} \Delta V p_{e,ijk}^{n}/(\gamma -1)$. To demonstrate discrete total energy conservation, it is sufficient to show that the total energy at the $n+1$ time-step is equal to that of the $n$th time-step, ie. $(K_i + K_e + W_B)^{n+1} - (K_i + K_e + W_B)^n = 0$.

The discrete change in total energy between the $n$ and $n+1$ time-step is given by the total ion kinetic energy, the magnetic energy, and the electron pressure equation~(\ref{discretePe}), as
\begin{align*}&\frac{(K_i + K_e +W_B)^{n+1} - (K_i + K_e + W_B)^n}{\Delta t} \\
&= \sum_{ijk}\Delta V \left[-\frac{\gamma}{\gamma - 1}\left(\boldsymbol{\nabla} \cdot \left[\boldsymbol{u}_e p_e\right]\right)_{ijk}^{n+1/2} - \left(\boldsymbol{\nabla}\cdot \boldsymbol{q}_e\right)_{ijk}^{n+1/2}\right] + \left[\boldsymbol{u}_{e,ijk}^{n+1/2}  - \boldsymbol{u}_{ijk}^{n+1/2} + \frac{\boldsymbol{j}_{ijk}^{n+1/2}}{en_{ijk}^{n+1/2}}\right]\cdot \left(\boldsymbol{\nabla} p_e\right)_{ijk}^{n+1/2}\\
&=0\end{align*}
by cancellation of terms, and using the definition of $\boldsymbol{u}_{e,ijk}^{n+1/2}$. The property $\sum_{ijk}\left(\boldsymbol{\nabla} \cdot \left[\boldsymbol{u}_e p_e\right]\right)_{ijk}^{n+1/2}=0$ has also been assumed here, which is satisfied for all conservative definitions of this advective flux term. We have not yet fixed the definition of this term, to allow for the flexible use of monotonicity preserving schemes in problems with strong flows, see below. In this derivation, the implicit midpoint scheme allows for the cancellation of field and particle contributions to the total energy change without temporal truncation error. In practice, energy is conserved down to the non-linear tolerance of the iterative method, as will be discussed below.

\section{Implementation}

\subsection{Non-linear solver strategy}\label{implementation}

The temporal and spatial discretization described above yields the set of algebraic equations~(\ref{discreteA}-\ref{umoment}), which can be written in the form $\boldsymbol{F}(\boldsymbol{x}^{n+1})=\boldsymbol{0}$ for the sought solution vector at the new time-step $\boldsymbol{x}^{n+1}=(\boldsymbol{x}_{p}^{n+1},\boldsymbol{v}_{p}^{n+1},\boldsymbol{A}_{ijk}^{n+1},p_{e,ijk}^{n+1})$. The system contains a number of non-linear implicit couplings, and to solve it in a scalable manner requires the use of iterative methods. Here we use the Jacobian-free Newton-Krylov (JFNK) method to converge the residual $\boldsymbol{F}$ to within a specified non-linear tolerance $\epsilon_t$. Our general implementation has been well documented in Refs.~\cite{knoll04,chacon08a}, but we give an overview here to motivate the discussion of our specific implementation below. The Newton linearization gives an equation for the solution at non-linear iteration $k+1$. 
\begin{equation}\label{NewtonLinearized}\frac{\partial \boldsymbol{F}}{\partial \boldsymbol{x}}\Bigg{|}^k \left(\boldsymbol{x}^{k+1}-\boldsymbol{x}^k\right) = -\boldsymbol{F}(\boldsymbol{x}^k),\end{equation}
which is iterated until $||\boldsymbol{F}(\boldsymbol{x}^k)||_2 < \epsilon_a + \epsilon_r ||\boldsymbol{F}(\boldsymbol{x}^0)||_2 = \epsilon_t$. Here $||\cdot||_2$ is the $L_2$-norm, $\epsilon_a = \sqrt{N}\times 10^{-15}$ is an absolute tolerance to avoid trying to converge below round-off (N is the total number of degrees of freedom), $\epsilon_r$ is the relative Newton convergence tolerance, and $\boldsymbol{F}(\boldsymbol{x}^0)$ is the initial residual. At every Newton iteration Eq.~(\ref{NewtonLinearized}) is solved using the flexible GMRES Krylov subspace method~\cite{saad93}. Constructing and storing the Jacobian matrix $\mathbb{J}=\partial \boldsymbol{F}/\partial \boldsymbol{x}$ can be extremely memory intensive and it unnecessary in practice. Instead, the Gateux derivative is used to calculate the needed matrix-vector products as 
\begin{equation}\frac{\partial \boldsymbol{F}}{\partial \boldsymbol{x}}\Bigg{|}^k \boldsymbol{v} = \frac{\boldsymbol{F}(\boldsymbol{x}^k + \epsilon \boldsymbol{v}) - \boldsymbol{F}(\boldsymbol{x}^k)}{\epsilon},\end{equation}
for a small and finite value of $\epsilon$. As in previous JFNK implementations, we use an inexact Newton's method where the convergence tolerance of the Krylov method varies for each Newton iteration depending on the size of the residual, see Refs.~\cite{chacon08a,chen11} for details. A key advantage of Krylov methods is that they can be preconditioned by solving the alternate systems $(\mathbb{P}^k)^{-1} \mathbb{J}^k \left(\boldsymbol{x}^{k+1}-\boldsymbol{x}^k\right) = -(\mathbb{P}^k)^{-1} \boldsymbol{F}^k$ (left preconditioning) or $\mathbb{J}^k (\mathbb{P}^k)^{-1} \mathbb{P}^k  \left(\boldsymbol{x}^{k+1}-\boldsymbol{x}^k\right) = -\boldsymbol{F}^k$ (right preconditioning). When $(\mathbb{P}^k)^{-1} \approx (\mathbb{J}^k)^{-1}$ the convergence properties of the Krylov iterations can be dramatically improved, while the solution of the Jacobian system is unchanged upon convergence. We leave the discussion of the preconditioning strategy for a future publication, and take the preconditioning matrix to be the identity matrix $\mathbb{P}^k = \mathbb{I}$ in this paper.


For the present discussion, we will use an electromagnetic field formulation taking the total solution vector to be $\boldsymbol{x}^{n+1} = (\boldsymbol{x}_{p}^{n+1},\boldsymbol{v}_{p}^{n+1},\boldsymbol{E}_{ijk}^{n+1/2},\boldsymbol{B}_{ijk}^{n+1},p_{e,ijk}^{n+1})$. In practice it is not necessary, and is indeed inefficient, to place the whole solution vector in the Newton residual as $\boldsymbol{F}(\boldsymbol{x}^{n+1})$. Non-linear elimination~(e.g. Ref.~\cite{chen11}) can be used to reduce the number of dependent variables in the residual from $\boldsymbol{F}(\boldsymbol{X}_1,\boldsymbol{X}_2)$ to $\boldsymbol{G}(\boldsymbol{X}_1)$ for new residual $\boldsymbol{G}$ when the dependent variable $\boldsymbol{X}_2$ can be written explicitly as $\boldsymbol{X}_2 = \boldsymbol{f}(\boldsymbol{X}_1)$. For the present set of equations there are two notable choices in the form of the residual, which differ in the convergence properties for momentum conservation and in algorithmic efficiency. The first is to form the Newton residual as $\boldsymbol{G}(\boldsymbol{v}_{p}^{n+1},\boldsymbol{B}_{ijk}^{n+1},p_{e,ijk}^{n+1})$, such that $(\boldsymbol{v}_{p}^{k+1},\boldsymbol{B}_{ijk}^{k+1},p_{e,ijk}^{k+1})$ is found from the $(k+1)$th iteration of~(\ref{NewtonLinearized}). The new position can be then directly calculated within the function evaluation for the next Newton iteration as $\boldsymbol{x}_{p}^{k+1}=\boldsymbol{x}_{p}^{k+1}(\boldsymbol{v}_{p}^{k+1})$, before evaluating Ohm's law statically from known quantities $\boldsymbol{E}_{ijk}^{k+1}=\boldsymbol{E}_{ijk}^{k+1}(\boldsymbol{x}_{p}^{k+1},\boldsymbol{v}_{p}^{k+1},\boldsymbol{B}_{ijk}^{k+1},p_{e,ijk}^{k+1})$. The primary advantage of this implementation is that the scatter and gather operations that are performed either side of the static evaluation of $\boldsymbol{E}_{ijk}^{k+1}$ are done with the same particle position $\boldsymbol{\boldsymbol{x}}_{p}^{k+1}$, in the same manner as in derivation~(\ref{momderiv}). This ensures momentum conservation at each Newton iteration to round-off, and thus gives a momentum error that is independent of the non-linear convergence tolerance. However, a major drawback of this method is due to the presence of the particle quantities $\boldsymbol{v}_{p}^{n+1}$ in the residual, which can be extremely memory intensive for typical problems where the number of particles greatly exceeds the number of grid points.


The second option is to solve for the Newton residual $\boldsymbol{H}(\boldsymbol{E}_{ijk}^{n+1/2},p_{e,ijk}^{n+1})=\boldsymbol{0}$ and solve for $\boldsymbol{B}_{ijk}^{n+1}=\boldsymbol{B}_{ijk}^{n+1}(\boldsymbol{E}_{ijk}^{n+1/2})$, $\boldsymbol{v}_{p}^{n+1}=\boldsymbol{v}_{p}^{n+1}(\boldsymbol{x}_{p}^{n+1},\boldsymbol{E}_{ijk}^{n+1/2},\boldsymbol{B}_{ijk}^{n+1})$, and $\boldsymbol{x}_{p}^{n+1}=\boldsymbol{x}_{p}^{n+1}(\boldsymbol{v}_{p}^{n+1})$ in the function evaluation. The major advantage of this method is the reduction in memory usage, and flexibility in the method for the particle push. Namely, a different time-step can be used to integrate the particle orbits (sub-cycling, see below), and the particle push can be optimized to run efficiently on the latest many-cores architectures using threading and vectorization. Due to the implicit coupling $\boldsymbol{v}_{p}^{n+1}(\boldsymbol{x}_{p}^{n+1}(\boldsymbol{v}_{p}^{n+1}))$, which arises from the interpolation of the fields to the particle position, an iterative method must be used to non-linearly converge the particles at each Newton iteration (see below). The momentum conservation is ensured down to the maximum of the particle orbit integration error and the outer non-linear convergence tolerance, which is demonstrated numerically in~\ref{implementationcomparison}. In this paper, we use the latter method modified for a potential formulation, i.e. with a residual $\boldsymbol{H}(\boldsymbol{A}_{ijk}^{n+1},p_{e,ijk}^{n+1})$. This allows us to easily implement the electromagnetic field and fluid equations in an existing Hall-MHD algorithm that uses a potential formulation.




\subsection{Sub-cycled and orbit-averaged particle push}\label{subcycleorbitav}


Taking the same time-step $\Delta t$ for both the particle orbit integration and the field advance can be inaccurate, e.g. Ref.~\cite{chen11}. For accuracy in the particle integration, it is necessary to resolve sufficiently the gyro-frequency $\Omega_{ci} \propto B$ that can vary strongly across the domain, depending upon the local magnetic field strength. However, solving for the electromagnetic fields at every time-step is unnecessary for the modeling of low-frequency phenomena where $\omega << \Omega_{ci}$. To address this, we use sub-stepping in the particle orbit integration. The sub-step for each particle $\Delta \tau^{\nu}_p$ is calculated using a local error estimator using the same method as Ref.~\cite{chen15}. Here the macro time-step for the field solve $\Delta t$ relates to the particle sub-steps as $\Delta t = \sum_{\nu = 0}^{N_{\nu p} -1} \Delta \tau^{\nu}_p$. It is assumed that the electromagnetic fields are static over the particle sub-steps, such that variations in the particle force occur only due to the change in particle position and velocity. Thus, the particle equations of motion are given by 
\begin{equation}\label{discreteXsc}\frac{\boldsymbol{x}_{p}^{\nu+1} - \boldsymbol{x}_{p}^\nu}{\Delta \tau_p^{\nu}} = \boldsymbol{v}_{p}^{\nu+1/2},\end{equation}
\begin{equation}\label{discreteVsc}\frac{\boldsymbol{v}_{p}^{\nu+1} - \boldsymbol{v}_{p}^\nu}{\Delta \tau_p^{\nu}} = \frac{q_p}{m_p} \left(\boldsymbol{E}_{p}^{*,\,\nu+1/2} + \boldsymbol{v}_{p}^{\nu+1/2} \times \boldsymbol{B}_{p}^{\nu+1/2}\right),\end{equation}
where 
\begin{equation}\label{scatterBsc}\boldsymbol{B}_{p}^{\nu+1/2} = \sum_{ijk} S\left(x_i - x_{p}^{\nu+1/2}\right) S\left(y_j - y_{p}^{\nu+1/2}\right) S\left(z_k - z_{p}^{\nu+1/2}\right)\left(\boldsymbol{\nabla}\times \boldsymbol{A}\right)_{ijk}^{n+1/2},\end{equation}
\begin{equation}\label{scatterEsc}\boldsymbol{E}_{p}^{*,\,\nu+1/2} = \sum_{ijk} S\left(x_i - x_{p}^{\nu+1/2}\right) S\left(y_j - y_{p}^{\nu+1/2}\right) S\left(z_k - z_{p}^{\nu+1/2}\right) \boldsymbol{E}^{*,\,n+1/2}_{ijk}.\end{equation}

To close the system of equations, the moments are computed via orbit-averaging~\cite{cohen82,chen11,sturdevant16b}, where a contribution to the moment is made at each particle sub-step. A significant advantage of this approach is in the reduction of numerical noise in the gathered moments since a phase-space position for each macro-particle is sampled multiple times within $\Delta t$. Our key consideration here is to preserve momentum and energy conservation for the orbit averaged formulation. It is straightforward to show that the above momentum-energy conservation holds when the moments are gathered as
\begin{equation}\label{nmomentorbitav}n_{ijk}^{n+1/2} = \frac{1}{\Delta V}\sum_p \frac{q_p}{e} \frac{1}{\Delta t} \sum_{\nu=0}^{N_{\nu p} - 1} S\left(x_i - x_{p}^{\nu+1/2}\right) S\left(y_j - y_{p}^{\nu+1/2}\right) S\left(z_k - z_{p}^{\nu+1/2}\right)\, \Delta \tau_p^{\nu},\end{equation}
\begin{equation}\label{umomentorbitav}\boldsymbol{u}_{ijk}^{n+1/2} = \frac{1}{n_{ijk}^{n+1/2} \,\Delta V}\sum_p \frac{q_p}{e} \frac{1}{\Delta t}  \sum_{\nu=0}^{N_{\nu p} - 1}  S\left(x_i - x_{p}^{\nu+1/2}\right) S\left(y_j - y_{p}^{\nu+1/2}\right) S\left(z_k - z_{p}^{\nu+1/2}\right) \boldsymbol{v}_{p}^{\nu+1/2}\, \Delta \tau_p^{\nu}.\end{equation}

Note that the plasma density $n_{ijk}^{n+1/2}$ is also computed using orbit averaging here. In the implicit energy conserving Vlasov-Darwin algorithm~\cite{chen11}, the density is calculated without orbit averaging at integer time-steps since it is only used in the charge conservation diagnostic. If the density is not orbit-averaged in the present case, then discrete momentum conservation is lost.

The sub-cycled particle update equations are solved with a Picard-iterated implicit Boris method~\cite{chen11,chen15,vu95}. The positions and momenta at Picard iteration $k$ are denoted by $\boldsymbol{x}_{p}^{k}, \boldsymbol{v}_{p}^{k}$, where $\boldsymbol{x}_{p}^{k=0}=\boldsymbol{x}_{p}^{\nu}$ and $\boldsymbol{v}_{p}^{k=0}=\boldsymbol{v}_{p}^{\nu}$ are the known values at the previous sub-step. The following steps are then iterated:
\begin{enumerate}
\item Evaluate $\boldsymbol{B}_{p}^{\nu+1/2,\,k}$, $\boldsymbol{E}_{p}^{\nu+1/2,\,k}$ as in Eqs.~(\ref{scatterBsc})-(\ref{scatterEsc}) using $\boldsymbol{x}_{p}^{\nu+1/2,\,k} = \tfrac{1}{2}(\boldsymbol{x}_{p}^\nu + \boldsymbol{x}_{p}^k)$,
\item $\hat{\boldsymbol{v}}^k = \boldsymbol{v}_{p}^{\nu} + \alpha^k \boldsymbol{E}_{p}^{\nu+1/2,\,k}$,
\item $\boldsymbol{v}_{p}^{\nu+1/2,\,k+1} = \left(\hat{\boldsymbol{v}}^k + \alpha^k \left[\hat{\boldsymbol{v}}^k\times \boldsymbol{B}_{p}^{\nu+1/2,\,k} + \alpha^k \left(\hat{\boldsymbol{v}}^k\cdot \boldsymbol{B}_{p}^{\nu+1/2,\,k}\right) \boldsymbol{B}_{p}^{\nu+1/2,\,k}\right]\right)/\left(1+\left(\alpha^k  \boldsymbol{B}_{p}^{\nu+1/2,\,k}\right)^2\right)$,
\item $\Delta \tau_p^{k+1} = \textrm{min}(\Delta \tau_p^k, dl_x/v_{px}^{\nu+1/2,\,k+1},dl_y/v_{py}^{\nu+1/2,\,k+1},dl_z/v_{pz}^{\nu+1/2,\,k+1})$,
\item $\boldsymbol{x}_{p}^{k+1} = \boldsymbol{x}_{p}^\nu + \boldsymbol{v}_{p}^{\nu+1/2,\,k+1} \Delta \tau_p^{k+1}$
\end{enumerate}
until the convergence condition $(|\boldsymbol{x}_{p}^{k+1}-\boldsymbol{x}_{p}^{k}| + |\boldsymbol{v}_{p}^{k+1}-\boldsymbol{v}_{p}^{k}|) < \epsilon_p$ is met, where a particle Picard tolerance of $\epsilon_p = 10^{-13}$ is typically employed. When this is satisfied the particle positions and momenta are updated for the next sub-step as $\boldsymbol{x}_{p}^{\nu+1} = \boldsymbol{x}_{p}^{k+1}$, $\boldsymbol{v}_{p}^{\nu+1} = 2\boldsymbol{v}_{p}^{\nu+1/2,\,k+1} - \boldsymbol{v}_{p}^\nu$. Here, $\alpha^k = \Delta \tau_p^k q_p/2m_p$, and $dl_x$, $dl_y$ and $dl_z$ are the distances in each direction to the closest cell boundary in the direction of motion. These are used to prevent particles from crossing cell boundaries within a sub-step, and so contribute to the density and velocity moments within every cell that they cross.

\subsection{Conservative smoothing}

To reduce discrete particle noise, a user-defined number of binomial smoothing passes can be applied to the field and moment quantities. For a cell-centered quantity $Q_{ijk}$, one pass of the smoothing operator is defined as $\textrm{SM}(Q_{ijk}) = \textrm{SM}_k(\textrm{SM}_j(\textrm{SM}_i(Q_{ijk})))$ where the smoothing in each direction is defined as
\begin{equation}\textrm{SM}_i(Q_{ijk}) = \frac{Q_{i-1jk} + 2Q_{ijk} + Q_{i+1jk}}{4}.\end{equation}

For suitable boundary conditions, e.g. periodic, the smoothing operator has the property $\sum_i A_{ijk} \textrm{SM}_i(B_{ijk}) = \sum_i \textrm{SM}_i(A_{ijk}) B_{ijk}$. It can be shown that global momentum and energy conservation is unaffected by such smoothing when the moments are gathered as
\begin{equation}n_{ijk}^{n+1/2} = \frac{1}{\Delta V}\textrm{SM}\left[\sum_p \frac{q_p}{e} S\left(x_i - x_{p}^{n+1/2}\right) S\left(y_j - y_{p}^{n+1/2}\right) S\left(z_k - z_{p}^{n+1/2}\right)\right],\end{equation}
\begin{equation}\boldsymbol{u}^{n+1/2}_{ijk} = \frac{1}{n_{ijk}^{n+1/2} \, \Delta V}\textrm{SM}\left[\sum_p \frac{q_p}{e} S\left(x_i - x_{p}^{n+1/2}\right) S\left(y_j - y_{p}^{n+1/2}\right) S\left(z_k - z_{p}^{n+1/2}\right) \boldsymbol{v}_{p}^{n+1/2}\right],\end{equation}
and the fields are scattered as
\begin{equation}\boldsymbol{B}_{p}^{n+1/2} = \sum_{ijk} S\left(x_i - x_{p}^{n+1/2}\right) S\left(y_j - y_{p}^{n+1/2}\right) S\left(z_k - z_{p}^{n+1/2}\right) \textrm{SM}\left[\left(\boldsymbol{\nabla}\times \boldsymbol{A}\right)_{ijk}^{n+1/2}\right],\end{equation}
\begin{equation}\label{scatterEsm}\boldsymbol{E}_{p}^{*,\,n+1/2} = \sum_{ijk} S\left(x_i - x_{p}^{n+1/2}\right) S\left(y_j - y_{p}^{n+1/2}\right) S\left(z_k - z_{p}^{n+1/2}\right) \textrm{SM}\left[\boldsymbol{E}^{*,\,n+1/2}_{ijk}\right].\end{equation}
This property is demonstrated numerically in Section~\ref{numerical}.

\subsection{Advection}\label{advection}

The conservative advection term in the electron pressure equation~(\ref{discretePe}), $\left(\boldsymbol{\nabla} \cdot \left[\boldsymbol{u}_e p_e\right]\right)_{ijk}^{n+1/2}$, gives no contribution to the global energy conservation error for suitably chosen boundary conditions. This allows some flexibility in the choice of its discretization. Here we allow a number of options, which are inherited from an existing MHD fluid framework~\cite{chacon04}. For linear problems we typically use the symmetric ZIP differencing~\cite{hirt68, chacon04}, which satisfies the discrete chain rule property and does not suffer from anti-diffusive truncation errors~\cite{chacon04}. However, this can be non-monotonic, particularly for non-linear problems with strong flows such as the reconnection problem discussed below. For such problems we instead use the Sharp and Monotonic Algorithm for Realistic Transport (SMART) algorithm~\cite{gaskell88}.

\subsection{Normalization}

The discrete model is implemented in normalized form using natural ion units. For a typical magnetic field strength $B_0$, density $n_0$, and ion mass $m_0$, velocities are normalized by the Alfv\'en speed $v_0 = v_{A0} = B_0/\sqrt{m_0 \mu_0 n_0}$, times by the inverse cyclotron period $t_0 = \Omega_{c0}^{-1} = m_0/eB_0$, and lengths by the ion skin depth $L_0 = d_{i0} = v_{A0}/\Omega_{c0}$. The different species of kinetic ions are distinguished by their normalized charge $Z_s = q_s/e$,  mass $M_s = m_s/m_0$, and density $N_s = n_{s0}/n_0$. If they are initialized with a Maxwellian distribution function with temperature $T_{s0}$, we define the thermal velocity as $v_{Ts0} = \sqrt{T_{s0}/m_s}$ and plasma-beta $\beta_{s0}/2 = M_s N_s (v_{Ts}/v_{A0})^2 = 2\mu_0 n_{s0}T_{s0}/B_0^2$. The electron sound speed is defined as $C_{s} = \sqrt{T_{e0}/m_0}$ and the ratio of ion to electron temperatures is $\tau_s = T_{s0}/T_{e0}$. 



\section{Numerical test of the finite-grid instability for a cold drifting beam}\label{coldbeam}

A major limitation of particle-in-cell algorithms to model the full Vlasov-Maxwell system is the need to resolve the electron Debye length to prevent unstable numerical heating. Such heating is typically demonstrated for the problem of a cold electron beam moving through a uniform periodic mesh at constant speed with a charge neutralizing ion background~\cite{birdsall91,okuda72}. The system is physically stable, however, spatial aliasing errors from particle-mesh interpolations can lead to a purely numerical instability that causes violation of momentum or energy conservation (or both), depending on the numerical scheme used~\cite{okuda72,brackbill16,huang16}. Finite-grid instabilities have also been studied using an explicit momentum conserving algorithm for the quasi-neutral hybrid model. Ref.~\cite{rambo95} presents a linear analysis of the hybrid finite grid instability with a discrete spatial grid for various shape functions, including nearest grid point (NGP) and linear interpolation/cloud-in-cell (CIC). Both the distribution function and the time advance is assumed to be continuous. It was found that a cold but finite ion temperature ($0<T_i/T_e < 1$) can make the beam unstable, where the precise instability threshold depends on the order of particle-mesh interpolation used. For example, instability occurs for $T_i/T_e <  1/12$ for NGP and $T_i/T_e < 1/25$ for CIC. Finite drift is further destabilizing and leads to growth rates $\gamma$ in the range $0.1<\gamma \Delta x/C_s<1$ for $T_i/T_e<20$ using NGP interpolation. 

Determining whether a PIC algorithm that simultaneously preserves both momentum and energy is absolutely stable with respect to finite-grid instabilities is a question left for future study, since it requires a dedicated analysis. Here, we present numerical evidence in support of this proposition using a particularly challenging case of a very cold ($T_i/T_e = 1/600$), drifting ($u_0/C_s = 0.1$) beam with NGP interpolation and with no smoothing. We compare the implicit conservative scheme with two other commonly used non-conservative schemes. The first scheme is a simple explicit leapfrog scheme that conserves momentum to numerical round-off, see~\ref{explicitmomentum} and Ref.~\cite{rambo95}. The second algorithm is an explicit predictor-corrector scheme, see~\ref{explicitpredcorr}, that uses an additional particle push to calculate the time-advanced electric field, but does not conserve momentum or energy due to truncation error. The problem set-up is chosen to be the same as for Figures 3 and 4 of Ref.~\cite{rambo95}; using a 1D-1V periodic domain of length $40\Delta x$, $50$ particles per cell, and time-step $\Delta t = 0.022 \Delta x/C_s$. 



\begin{figure}
\begin{center}
\includegraphics[width=0.7\textwidth]{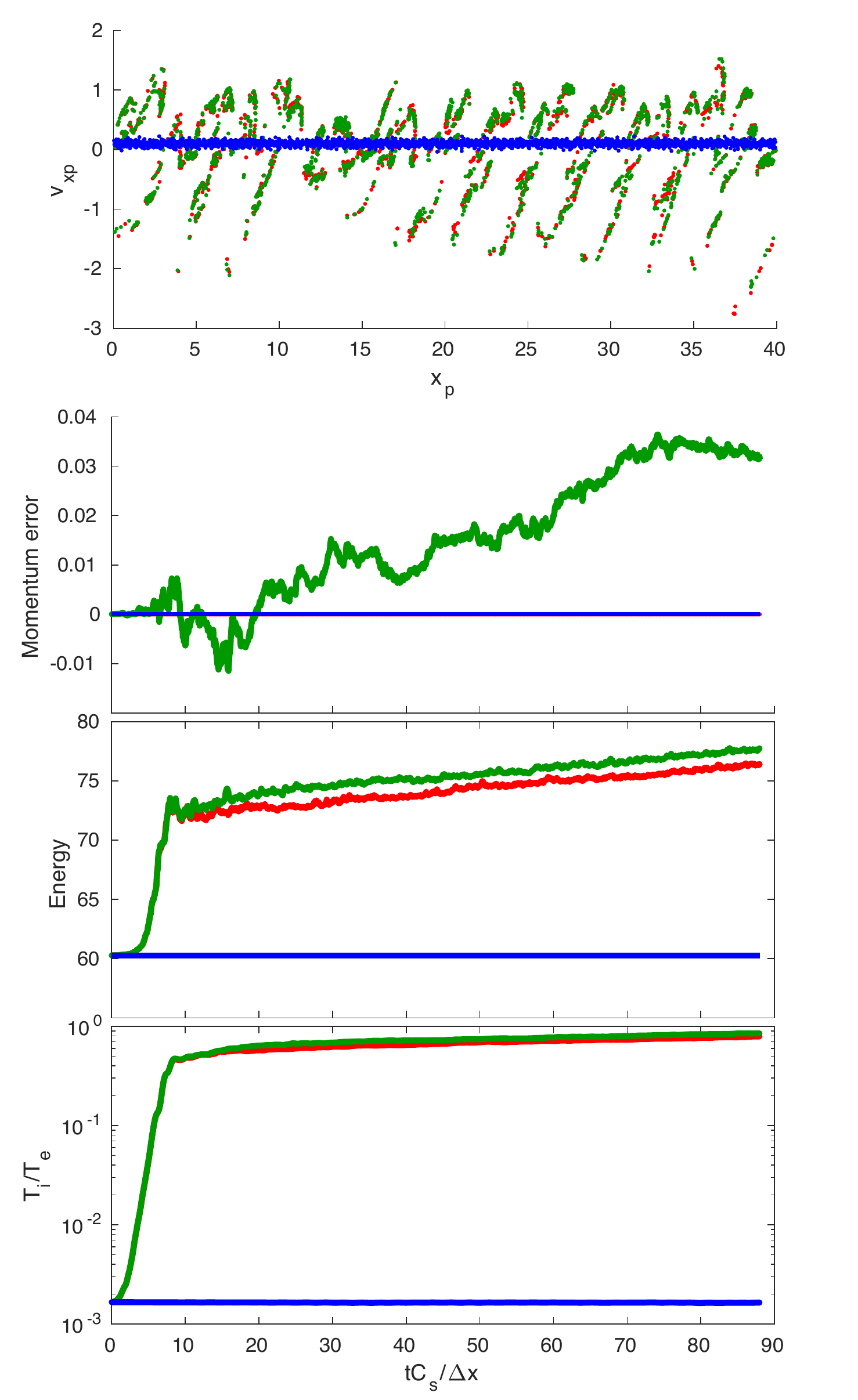}
\caption{\label{coldbeamfig}Comparison study between an explicit leapfrog-based momentum-conserving scheme (red, see~\ref{explicitmomentum}), a predictor-corrector method (green, see~\ref{explicitpredcorr}), and the new implicit conservative algorithm (blue). Top panel: Particle phase space distribution $(x_p,v_{xp})$ at $t=8.8 \Delta x/C_s$ (400 steps). Second panel: Change in total momentum vs time (the red and blue traces overlap). Third panel: Total energies (ion kinetic + electron thermal) vs time. Bottom panel: Ion-to-electron temperature ratio vs time on a logarithmic scale, with initial $T_i/T_e = 1/600$.}
\end{center}
\end{figure}

Figure~\ref{coldbeamfig} shows the numerical results for the cold-beam problem for the three algorithms. The top panel shows the particle phase-space distribution $(x_p,v_{xp})$ after 400 time-steps ($t=8.8 \Delta x/C_s$) for the momentum-conserving leapfrog algorithm (red), the explicit predictor-corrector scheme (green), and the new implicit conservative scheme (blue). The phase space distributions for the first two (explicit) algorithms show characteristic fluctuations of the finite-grid instability, while the implicit algorithm appears stable. The effects of these fluctuations on the momentum and energy conservation are shown in the second and third panels in Fig.~\ref{coldbeamfig}. In the second panel, the momentum error, ($P_{x}^n - P_{x}^0$), remains of the order of $10^{-14}$ for the two schemes that feature discrete momentum conservation, while there are significant errors in momentum conservation ($\approx 0.03$) for the predictor-corrector scheme. The total energy grows by $20-30\%$ for both of the explicit schemes, in contrast to the new implicit scheme that conserves energy down to $\approx 10^{-13}$ (for relative non-linear tolerance $\epsilon_r = 10^{-14}$). We note that discrete global energy conservation in the implicit conservative scheme does not preclude conversion of energy between the ion and electron kinetic energies. Indeed, spatial aliasing errors are present in all particle-in-cell schemes with a spatial grid, and can cause small variations in the energies of each ion particle. It is thus instructive to also consider the ion-to-electron temperature ratio, which is one measure of this energy partition between the kinetic and fluid components. The bottom panel shows $T_i/T_e$ for the three schemes on a logarithmic axis. Here, the two explicit schemes exhibit linear growth which saturates at $T_i/T_e = 0.8$ for the momentum-conserving leapfrog-based algorithm and $T_i/T_e = 0.85$ for the predictor-corrector scheme. The saturation level differs by a small amount from the value of $T_i/T_e = 0.625$ mentioned in Ref.~\cite{rambo95}, and we have verified that these differences are due to the use of an adiabatic, rather than isothermal, equation of state and our choice of discretization of the electric field (the explicit momentum-conserving algorithm in~\ref{explicitmomentum} is equivalent to Ref.~\cite{rambo95} in the linear regime). There is a small decrease ($<1\%$) in the ratio of $T_i/T_e$ in the implicit conservative scheme which is non-monotonic over the timescale plotted. However, we see no sign of the unstable growth of $T_i/T_e$ due to the finite grid instability for $10^6$ time-steps ($t=22000 \Delta x/C_s$), which is as long as we have run the simulation.

\section{Numerical Verification}\label{numerical}

In the remainder of the paper, we present numerical examples that are intended to verify the correctness of our implementation, and demonstrate the conservation properties as described above. For iterative implicit methods, such as the one discussed here, the cost of the algorithm strongly depends on the effectiveness of the preconditioner used. This is particularly the case for implicit particle methods, where the number of particle pushes is proportional to the number of function evaluations, and the cost of applying a grid-based preconditioner is small compared to that of each particle push. A thorough discussion of the preconditioning strategy and the cost to reach a given non-linear convergence is left for a separate publication. Here, we will typically converge the non-linear tolerance close to numerical round-off error in the following examples, to demonstrate the absence of any spatial or temporal truncation errors in the momentum and energy conservation. The choice of the macro time-step $\Delta t$ is not limited by stability considerations, but we typically choose it small to reduce stiffness in the absence of a preconditioner. In practice, the conservation properties are only ensured down to the chosen non-linear tolerance, which is demonstrated in~\ref{implementationcomparison}. We do not use heat conductivity for any of the problems $(\kappa_e=0$), and the resistive and hyper-resistive dissipation is set to zero in all examples except for the non-linear reconnection problem. 

\subsection{Landau damped ion acoustic wave (1D-1V electrostatic)}\label{iawproblem}

\begin{center}
\begin{figure}
\includegraphics[width=\textwidth]{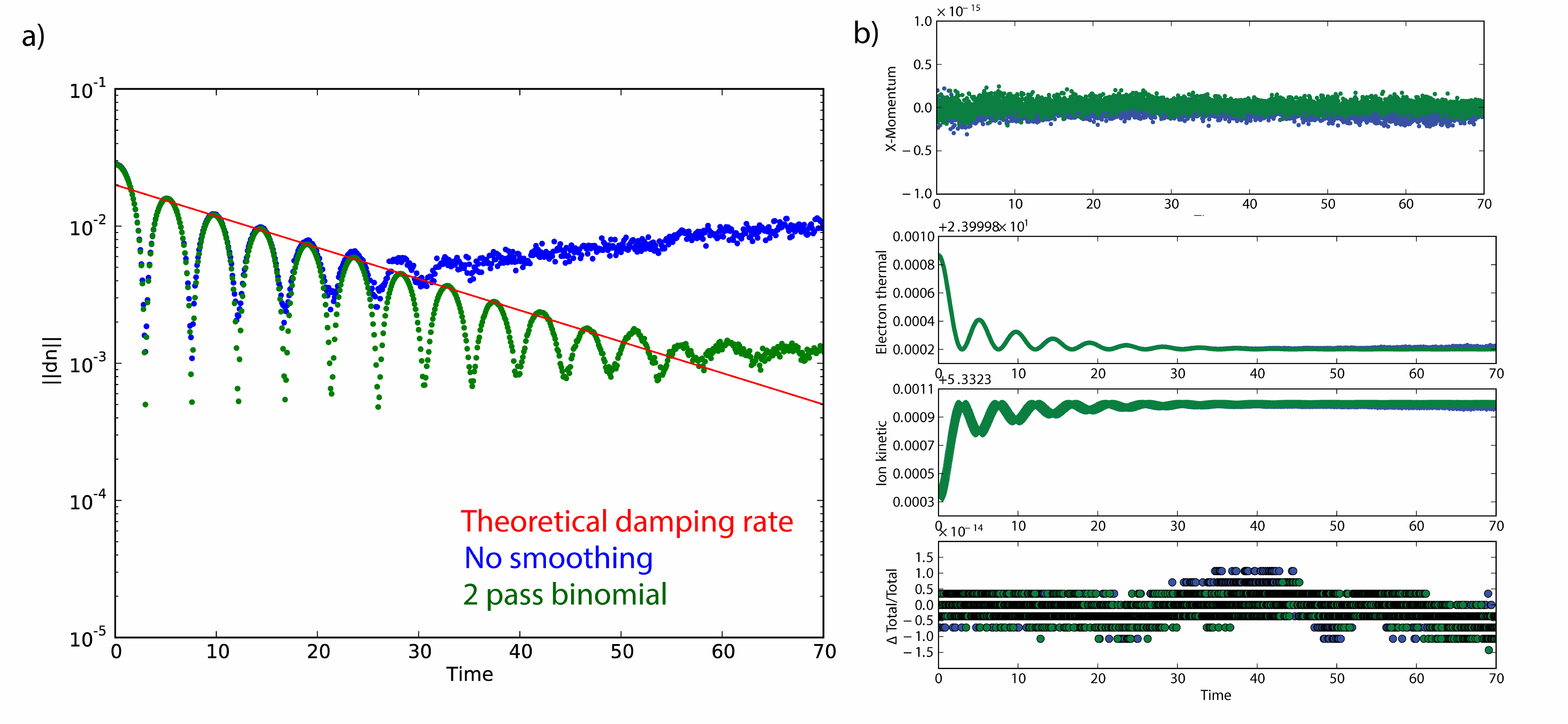}
\caption{\label{iawtest}a) L$^2$-norm of the density perturbation against time for the Landau-damped ion acoustic wave problem. b) Traces of the momentum, electron thermal energy, ion kinetic energy and relative total energy change for the same simulations. The green trace is for a run with two passes of binomial smoothing, while the blue is without smoothing.}
\end{figure}
\end{center}

The first numerical example we consider is the 1D-1V electrostatic ion acoustic wave (IAW). Due to ion kinetic effects, the wave is damped via ion Landau damping. We initialize the problem with a single-ion species Maxwellian plasma with ion charge $Z_i = 1$, mass $M_i = 1$, thermal velocity $v_{Ti0} = \sqrt{1/3}$, and quasi-neutral density $N_i = 1+0.01\sin{(k x)}$ with wavenumber $k=\pi/8$ and domain length $2\pi/k$. The electron fluid has $\gamma=5/3$ and we use a temperature ratio of $\tau = 0.2$, which gives an electron sound speed of $C_s = \sqrt{5/3}$. The dispersion relation is given by $Z'(\xi) = 2\tau$, where $Z$ is the plasma dispersion function and $\xi = (\omega + i \gamma)/k\sqrt{2}v_{Ti0}$ is the normalized complex frequency.  Figure~\ref{iawtest}a) shows the $L^2$-norm of the density perturbation $||dn||=||N_i -1||$ from two simulations with $96$ uniform cells, and $50,000$ particles per cell that are sampled from the ion distribution function using a low-discrepancy Hammersley sequence (quasi-quiet start). The simulations are performed with a time-step of $\Delta t = 0.02$ and are converged to a relative Newton tolerance of $\epsilon_r=10^{-12}$ using ion sub-cycling and orbit averaging with a Picard tolerance of $\epsilon_p = 10^{-13}$. This tight tolerance typically requires 6 non-linear iterations and 6 linear iterations per time-step. The green curve shows the result with two applications of binomial smoothing to the fields and moments as described above, whereas the blue curve is the same run without any smoothing. In both cases, good agreement is found with linear theory for the real frequency $\omega=0.6827$ ($2.13\,k\sqrt{2}v_{Ti0}$)  and damping rate $\gamma=-0.05265$ ($-0.164\,k\sqrt{2}v_{Ti0}$), where the latter is indicated by the red line in Fig.~\ref{iawtest}a). The initial quasi-quiet start degrades over time due to the accumulation of numerical error, until the noise floor overwhelms the damped signal. Since there is significant separation between the grid resolution and the wavelength of the ion acoustic wave, binomial smoothing can be safely applied in this problem to reduce the impact of discrete particle noise. As shown in Fig.~\ref{iawtest}a), two passes of binomial smoothing significantly improves the solution at late time.

Figure~\ref{iawtest}b) shows the momentum error $(P_x^n-P_x^0)$, the electron thermal energy $K_e^n$, the ion kinetic energy $K_i^n$, and the relative total energy error $[(K_i + K_e)^n - (K_i + K_e)^0]/(K_i + K_e)^0$ for the two cases. The chosen relative tolerance gives errors on the level of double precision round-off: $~10^{-15}-10^{-16}$ for the momentum error, and $~10^{-14}$ for the energy error. These momentum and energy errors are unaffected by binomial smoothing, as discussed above, although there are some minor differences in the energy partition ($K_e^n$ vs $K_i^n$) between the two runs. 

\subsection{Proton Cyclotron Anisotropy Instability 1D-3D electromagnetic}

The Proton Cyclotron Anisotropy Instability (PCAI) is a fully electromagnetic instability driven by anisotropy in the proton temperature with respect to the background magnetic field with $T_{p\perp}/T_{p\parallel} > 1$. The mode has finite real frequency with maximum growth at $\boldsymbol{k}\times \boldsymbol{B}_0 = \boldsymbol{0}$ for wave vector $\boldsymbol{k}$ and background field $\boldsymbol{B}_0$. We simulate the problem in 1D-3V, where both the wave vector $\boldsymbol{k}$ and the background field $\boldsymbol{B}_0$ are parallel to the $x$-axis. The first case considered is a pure hydrogen plasma with $Z_p=M_p=N_p=1$ drawn from a uniform bi-Maxwellian distribution with $\beta_{p\parallel} = 1$, and $T_{p\perp}/T_{p\parallel} = 3$. The second case uses the same parameters but is comprised of $20\%$ doubly ionized helium ions (alpha particles), with $Z_\alpha = 2$, $M_\alpha = 4$, $T_{\alpha \parallel}/T_{p \parallel} = 2$, $T_{\alpha \perp}/T_{\alpha \parallel} = T_{p\perp}/T_{p\parallel}$, $N_\alpha/N_p = 0.2$ and $Z_\alpha N_\alpha+Z_pN_p = 1$. In both cases, the electron fluid is initialized with temperature ratio $\tau_p = 1$ and $\gamma = 5/3$. The simulation domain has a length of $10.5$, using $64$ cells with $50,000$ protons and $20,000$ helium ions per cell. The time-step is $\Delta t = 0.01$ and relative Newton tolerance $\epsilon_r = 10^{-12}$, which requires about $5$ non-linear and $6$ linear iterations per time-step. We use a quasi-quiet start and apply two passes of binomial smoothing.

\begin{figure}
\begin{center}
\includegraphics[width=0.5\textwidth]{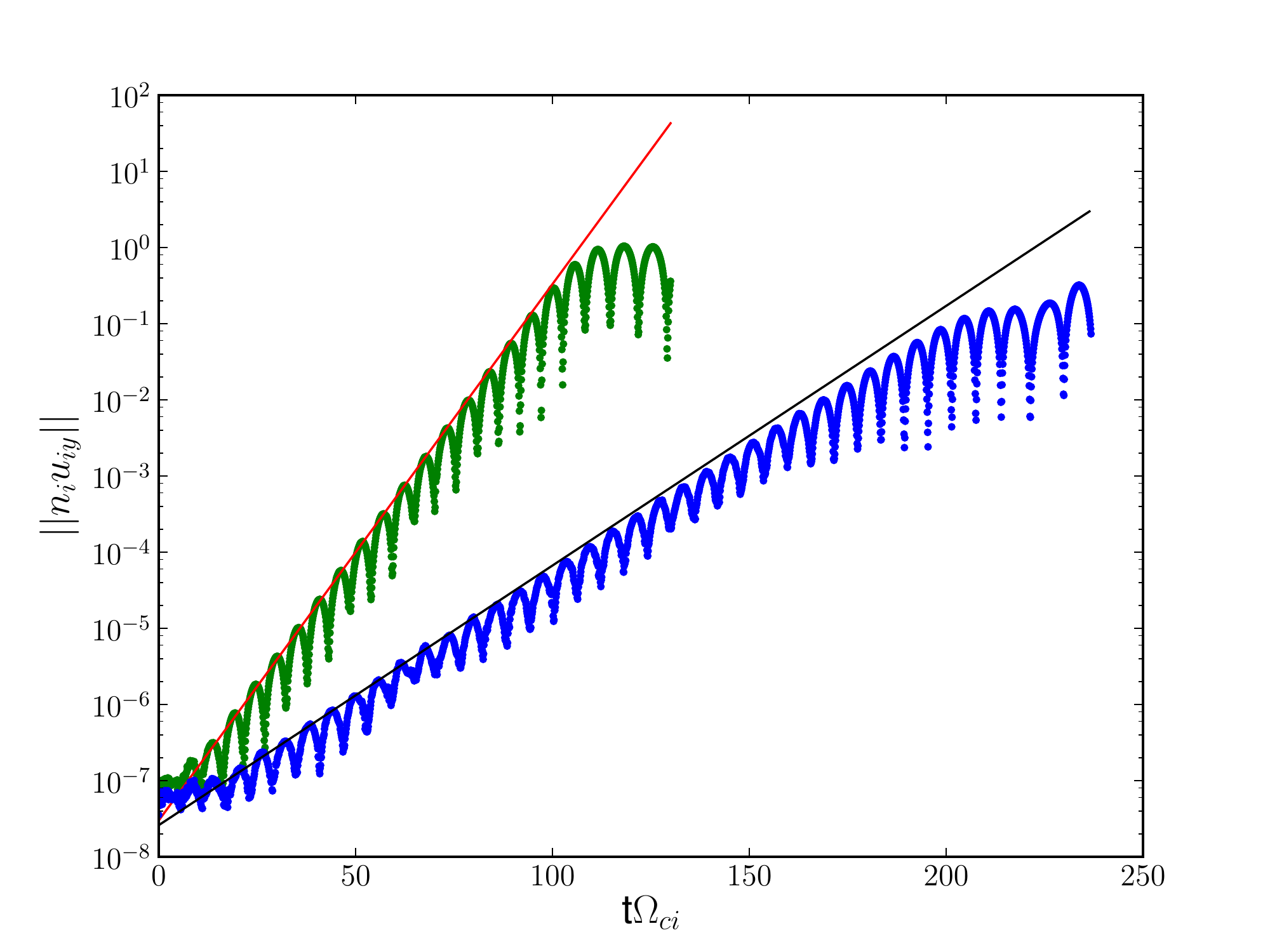}
\caption{\label{icaigrowth}L$^2$-norm of the transverse ion bulk momentum perturbation $||n_iu_{iy}||$ for the Proton Cyclotron Anisotropy Instability with parameters given in the text. Shown are results from a pure hydrogen plasma (green) with theoretical growth rate from linear theory (red), and for a plasma with a 20\% ion composition of alpha particles (blue) with theoretical growth rate (black).}
\end{center}
\end{figure}

Figure~\ref{icaigrowth} shows the L$^2$-norm of the transverse bulk ion momentum $||n_i u_{iy}||$ vs time. For the pure hydrogen plasma (green) there is good agreement between the simulation and the theoretical linear growth rate (red, $\gamma=0.162$ ~\cite{gary93book,told16}) over approximately six decades of growth in the perturbation amplitude. Studies of pressure anisotropy driven instabilities in the magnetosheath have found that the linear growth rate of the PCAI can reduce in the presence of doubly ionized helium ions. Theory has shown that the reduction in the growth rate can lead to a change from cyclotron to mirror-like fluctuations in some regions of the Earth's magnetosheath~\cite{price86,gary93}, in agreement with \textit{in-situ} spacecraft data, e.g. Ref.~\cite{hubert89}. For the parameters used here, the linear growth rate is reduced to $\gamma = 0.0785$ (black fit) and there is good agreement with the fully non-linear simulation that includes the two ion species (blue). For a relative Newton tolerance of $\epsilon_r = 10^{-12}$ used here, the momentum and relative total energy errors (not shown) are on the order of $10^{-14}$ and $10^{-12}$, respectively, for both of the non-linear simulations. 







\subsection{Non-linear magnetic reconnection}\label{reconnection}

The final numerical example is a 2D-3V simulation of magnetic reconnection. Parameters are chosen to reproduce the GEM challenge reconnection problem~\cite{birn01}, but with two current layers and double periodic boundary conditions. The simulation domain is a square box of dimensions $Lx = 25.6$, $Ly= 25.6$ with $384\times 384$ cells and an average of $320$ particles/cell. The time-step is $\Delta t = 0.01$ and we use a relative Newton tolerance $\epsilon_r = 10^{-13}$. A single ion species with $Z_i=M_i=1$ is initialized from an isotropic Maxwellian distribution function with $\beta_{i0} = 5/6$ (the total ion+electron $\beta = 1$). The initial density is non-uniform with function
\begin{equation}N_i = \frac{1}{\cosh^2(x/\lambda)} + \frac{1}{\cosh^2((x- Lx/2)/\lambda)} + \frac{1}{\cosh^2((x-Lx)/\lambda)} + 0.2.\end{equation}
The simulation is initiated with a large sinusoidal perturbation with amplitude $0.1B_0$. The electron fluid is initialized with $\gamma=5/3$ and a temperature ratio of $\tau=5$. In contrast to the previous numerical examples, finite dissipation is included here to break the frozen in condition and allow reconnection. We use a uniform resistivity $\eta = 5\times 10^{-4}$ and hyper-resistivity $\eta_H = 2\times 10^{-4}$ to break the frozen-in condition and allow reconnection.

\begin{center}
\begin{figure}
\includegraphics[width=\textwidth]{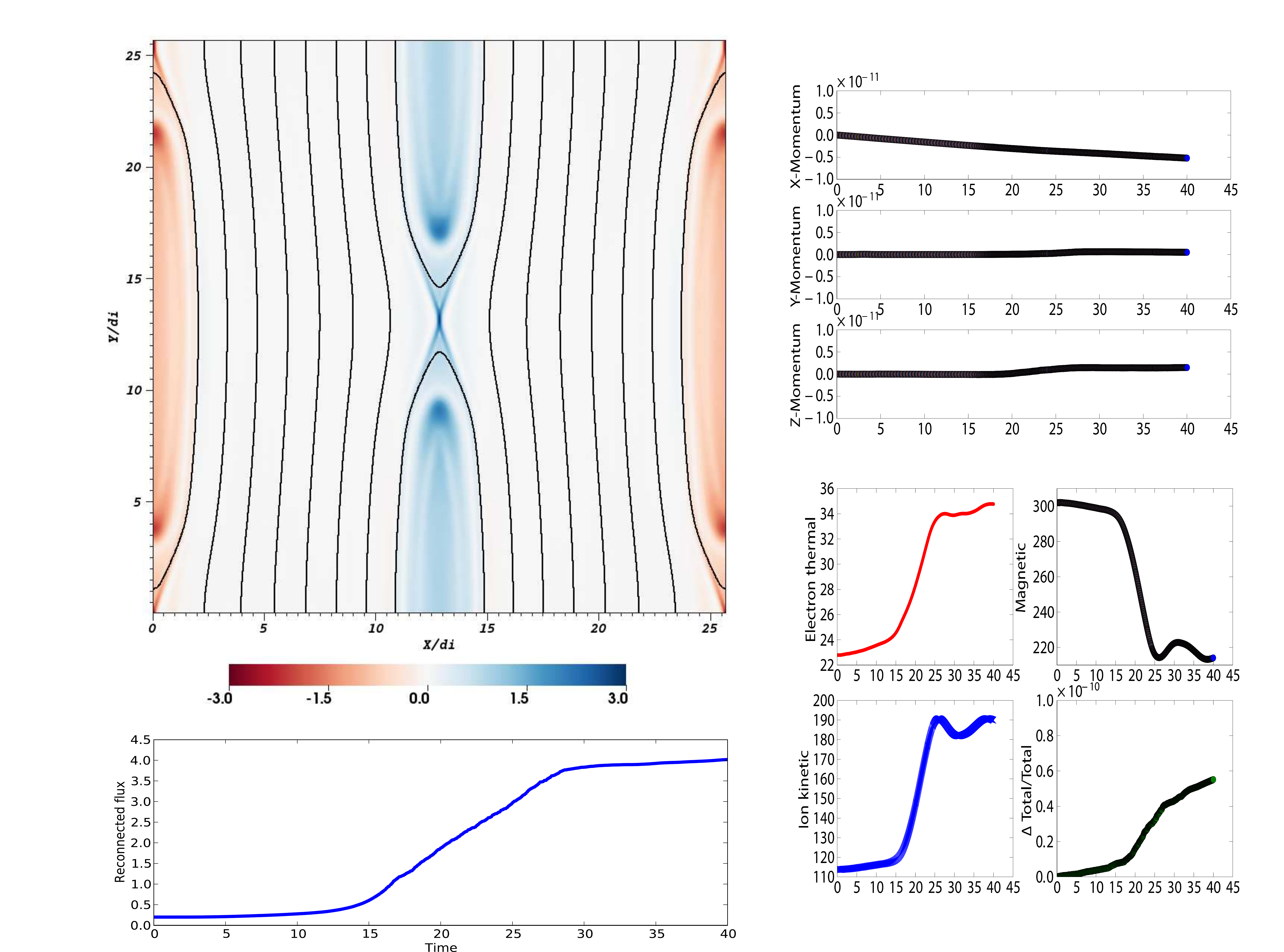}
\caption{\label{gemdp}Results from a double periodic version of the GEM challenge magnetic reconnection problem. Top left: Current density $j_z$ and contours of the magnetic potential $A_z$. Bottom left: Reconnected flux vs time. Top right: Momentum errors in each direction vs time. Bottom right: Traces of the electron thermal energy, magnetic energy, ion kinetic energy and the relative total energy error vs time. }
\end{figure}
\end{center}

Figure~\ref{gemdp} shows the results and conservation properties from the simulation. The top left panel shows the current density $j_z$ and contours of the vector potential $A_z$ at $t=20$ during the non-linear phase. As is typical for the hybrid model, e.g. Ref.~\cite{karimabadi04}, the out-of-plane current $j_z$ is localized around the magnetic X-point. We note that an anisotropic closure for the electron pressure can capture elongated current sheets within the hybrid model~\cite{le16}, which are more similar to those found in full kinetic simulations, particularly in the low-$\beta$ regime. Associated with these magnetic and current structures are strong outflows, which can lead to unphysical oscillations when using ZIP differencing to treat the conservative advection term in the electron pressure equation. As discussed in Section~\ref{advection}, we find that using the SMART algorithm for this advection term preserves monotonicity and removes such oscillations. The bottom left panel in Fig.~\ref{gemdp} shows the reconnected flux over time. Both the reconnection rate, $(\partial_t A_z)|_X \approx 0.25$ measured at the X-point, and the saturation level agree with those reported in Ref.~\cite{birn01} for a single current layer set-up. The momentum and energy conservation results are shown in the right panel. In this case the momentum error in each component direction is on the order of $10^{-10} - 10^{-11}$ and the relative total energy error is of the order $10^{-10}$. Due to the significant stiffness from the whistler wave dispersion and the high-order hyper-resistive dissipation, the simulation typically requires 12 nonlinear and 200 linear iterations per time-step to converge to this tight tolerance. This motivates the development of an optimal preconditioning strategy for the hybrid model, which we will pursue in future work.

\section{Summary}

A novel implicit, non-linear, particle-in-cell scheme has been formulated for the electromagnetic quasi-neutral hybrid model featuring global conservation of mass, momentum, and energy. The scheme uses the implicit-midpoint method for time advance, with field and fluid quantities defined at cell centers and using central differences for the numerical derivatives. Momentum and energy conservation is ensured by exact cancellation between particle moment quantities with terms in Ohm's law and the electron pressure equation that that exist on the spatial grid. These conservation properties hold for arbitrary-order shape functions (provided that the same shape function is used throughout), and are also preserved when sub-cycling the particle advance using a Picard-iterated implicit Boris push, collecting orbit-averaged moments, and applying conservative smoothing. The discrete equations are implemented within an iterative Jacobian-free Newton Krylov solver, and we have compared two possible forms of the Newton residual in terms of the convergence of the momentum conservation error, and the potential for optimization of the particle advance routines. 

The favorable stability properties of the algorithm are demonstrated for the problem of a cold beam moving through a uniform periodic mesh~\cite{rambo95}. There is no indication of the finite grid instability for the new conservative scheme, for a challenging set of parameters in which two commonly used non-conservative schemes are unstable. Future work will explore these stability properties in more detail, comparing with linear theory for the instability that includes the effects of a spatial grid. 

The correctness of implementation, and the conservation properties, are then demonstrated for a number of test problems including the Landau damping of the ion acoustic wave, the Proton Cyclotron Anisotropy Instability (PCAI), and non-linear magnetic reconnection. For the non-linear problem, it was found that oscillations due to sharp gradients and strong flows were eliminated by the use of the SMART scheme~\cite{gaskell88} for the conservative advection term in the electron pressure equation. Additional future work will explore the possibilities of discrete particle noise reduction via delta-$f$ methods~\cite{parker93} or remapping~\cite{wang11}, and the implementation of an optimal preconditioner for the electron model.

\section*{Acknowledgements}
This work is supported by the Applied Scientific Computing Research (ASCR) program of the U.S. Department of Energy, and used resources provided by the Los Alamos National Laboratory Institutional Computing Program, which is supported by the U.S. Department of Energy National Nuclear Security Administration under Contract No. DE-AC52-06NA25396. A.S. would like to thank William Taitano for helpful discussions.

\appendix

\section{Hyper-resistivity}\label{hyperresistivity}

In a typical symmetric set-up, where the location of the magnetic X-point coincides with the stagnation point for the ion and electron flows, some non-ideal term is required to balance the reconnection electric field. Using only resistivity is often inadequate, because it is unable to support sub-layers in the electron current which thin over time~\cite{zocco09,stanier15b}. Here, we include the flexibility to use higher-order hyper-resistive friction. This is able to support the electron layer even when reconnection becomes quasi-steady~\cite{stanier15b}, and also has an added benefit that it can set a dissipation scale for dispersive whistler waves that can otherwise lead to grid-scale noise in non-linear simulations, e.g. Ref.~\cite{chacon03}. 

We treat the hyper-resistivity as a friction term, which is added to the right hand side of Eq.~(\ref{discreteA}) but is not included in $\boldsymbol{E}^{*,\,n+1/2}_{ijk}$. We discretize the term using a compact stencil as
\begin{equation} - \eta_H \left(\nabla^2 \boldsymbol{j}\right)_{ijk}^{n+1/2} = -\eta_H \left[d_x^2 \boldsymbol{j} + d_y^2 \boldsymbol{j} + d_z^2 \boldsymbol{j}\right]^{n+1/2},\end{equation}
where the second derivative is defined as $(d_x^2 \boldsymbol{j})^{n+1/2} = (\boldsymbol{j}_{i+1jk}^{n+1/2} - 2 \boldsymbol{j}_{ijk}^{n+1/2} + \boldsymbol{j}_{i-1jk}^{n+1/2})/\Delta x^2$.

The term has no effect on momentum conservation, but, since it causes dissipation of magnetic energy, it is necessary to add a corresponding hyper-resistive viscous heating term in the electron pressure equation for total energy conservation. In discrete form, global energy is conserved for suitable boundary conditions when the heating term is defined as
\begin{equation}\eta_H \left(\boldsymbol{\nabla}\boldsymbol{j}:\boldsymbol{\nabla}\boldsymbol{j}\right)_{ijk}^{n+1/2} = \eta_H \sum_{\alpha=x,y,z} \left[\frac{(d_xj_\alpha)^2_{i+1/2jk} + (d_xj_\alpha)^2_{i-1/2jk}}{2} + \frac{(d_yj_\alpha)^2_{ij+1/2k} + (d_yj_\alpha)^2_{ij-1/2k}}{2} + \frac{(d_zj_\alpha)^2_{ijk+1/2} + (d_zj_\alpha)^2_{ijk-1/2}}{2}\right]^{n+1/2},\end{equation}
where the spatial derivatives are defined as $(d_x j_\alpha)^{n+1/2}_{i+1/2jk} =  \left[(j_\alpha)_{i+1jk} - (j_\alpha)_{ijk}\right]^{n+1/2}/\Delta x$. The conservation of global energy is demonstrated numerically for the reconnection problem in Section~\ref{reconnection}.

\section{Locally-mass-conserving measures of density and momentum}\label{massconsv}

For the commonly used B-spline shape functions, Refs.~\cite{chen11,chen15} presented a mass conserving scheme using quadratic splines for the density and products of linear and quadratic shape functions for the fluxes, where linear interpolation was used in the direction of each component. In particular, in 3D it can be shown that the definitions
\begin{equation}n_{ijk}^n = \frac{1}{\Delta V}\sum_p S_2\left(x_i - x_p^{n}\right)S_2\left(y_j - y_p^{n}\right)S_2\left(z_k - z_p^{n}\right),\end{equation}
\begin{equation}(nu_x)_{i+1/2jk}^{n+1/2} = \frac{1}{\Delta V}\sum_p v_{xp}^{n+1/2} S_1 \left(x_{i+1/2} - x_p^{n+1/2}\right) \mathbb{S}_{22}^{n+1/2}(y_j - y_p, z_k - z_p),\end{equation}
\begin{equation}(nu_y)_{ij+1/2k}^{n+1/2} = \frac{1}{\Delta V}\sum_p v_{yp}^{n+1/2} S_1 \left(y_{j+1/2} - y_p^{n+1/2}\right) \mathbb{S}_{22}^{n+1/2}(z_k - z_p, x_i - x_p),\end{equation}
\begin{equation}(nu_z)_{ijk+1/2}^{n+1/2} = \frac{1}{\Delta V}\sum_p v_{zp}^{n+1/2} S_1 \left(z_{k+1/2} - z_p^{n+1/2}\right) \mathbb{S}_{22}^{n+1/2}(x_i - x_p, y_j - y_p),\end{equation}
satisfy a local mass continuity equation on a staggered grid
\begin{equation}\label{localchargeconsv}\frac{n_{ijk}^{n+1}-n_{ijk}^n}{\Delta t} + \frac{(nu_x)_{i+1/2jk}^{n+1/2}-(nu_x)_{i-1/2jk}^{n+1/2}}{\Delta x} + \frac{(nu_y)_{ij+1/2k}^{n+1/2}-(nu_y)_{ij-1/2k}^{n+1/2}}{\Delta y} + \frac{(nu_z)_{ijk+1/2}^{n+1/2}-(nu_z)_{ijk-1/2}^{n+1/2}}{\Delta z} =0,\end{equation}
where  e.g. $(i+1/2jk)$ is the positive cell face of cell $(ijk)$,
\begin{equation}\mathbb{S}_{22}^{n+1/2} (\bar{y},\bar{z}) \equiv \tfrac{1}{6}\left[2 S_2(\bar{y}^{n+1})S_2(\bar{z}^{n+1}) + S_2(\bar{y}^n)S_2(\bar{z}^{n+1}) + S_2(\bar{y}^{n+1})S_2(\bar{z}^{n}) + 2S_2(\bar{y}^{n})S_2(\bar{z}^{n})\right].\end{equation}
Here, the shape functions are defined as
\begin{equation}S_1(x_{i-1/2} - x_p) = 1 - \frac{x_p - x_{i-1/2}}{\Delta x},\end{equation}
\begin{equation}S_1(x_{i+1/2} - x_p) = \frac{x_p - x_{i-1/2}}{\Delta x},\end{equation}
\begin{equation}\label{s2shape1}S_2(x_i - x_p) = \frac{3}{4} - \left(\frac{x_p - x_i}{\Delta x}\right)^2,\end{equation}
\begin{equation}\label{s2shape2}S_2(x_{i \pm 1}-x_p) = \frac{1}{2}\left(\frac{1}{2} \pm \frac{x_p - x_i}{\Delta x}\right)^2.\end{equation}
This method requires that the particle does not cross a cell boundary between $n\Delta t \leq t \leq (n+1)\Delta t$. Refs.~\cite{chen11,chen15} ensure this within a sub-cycled and orbit-averaged scheme by stopping the sub-step when particles hit cell boundaries. Since we use the same particle advance, also stopping particles as they cross cell boundaries, we are able to measure these locally-mass-conserving measures from the particle distribution at any time-step (in post-processing). However, we choose not to use such measures, as they do not allow momentum conservation for the hybrid model.

\section{Non-linear convergence of momentum and energy errors}\label{implementationcomparison}

\begin{center}
\begin{figure}
\includegraphics[width=\textwidth]{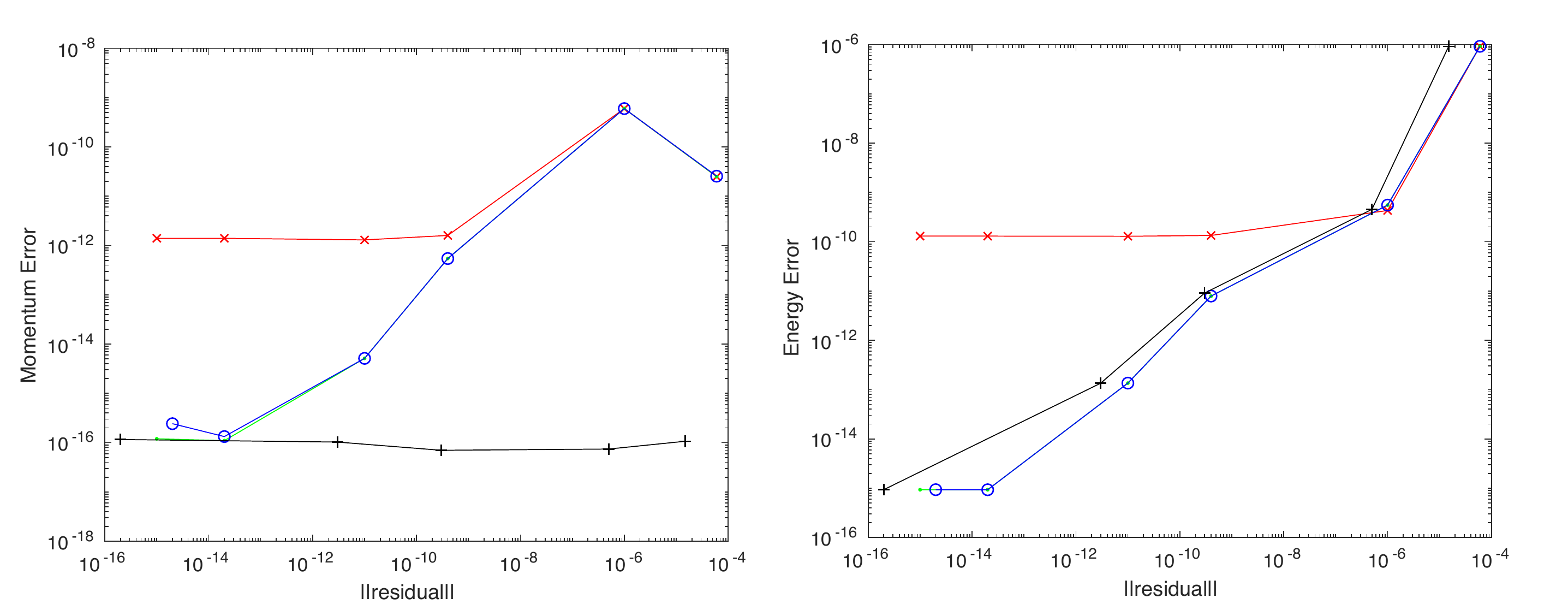}
\caption{\label{iaw-errors-vs-residual}Left: Momentum error vs. L$^2$-norm of the residual. Right: Relative energy error vs. residual norm. Results are from the first implementation with Newton residual $\boldsymbol{G}(\boldsymbol{v}_{p}^{n+1},p_{e,ijk}^{n+1})$ (black +), and the second implementation with Newton residual $\boldsymbol{H}(\boldsymbol{E}_{ijk}^{n+1/2},p_{e,ijk}^{n+1})$ and with particle orbit Picard-iteration tolerances $\epsilon_p = 10^{-4}$ (red x), $10^{-8}$ (green $\cdot$), and $10^{-12}$ (blue $\odot$). }
\end{figure}
\end{center}

Figure~\ref{iaw-errors-vs-residual} shows the convergence properties of the momentum and energy errors for the two different implementations of the discrete model, discussed in Section~\ref{implementation}, for the ion acoustic wave (IAW) problem. The problem set-up is the same is described in Section~\ref{iawproblem}. Since there is zero magnetic field for the IAW problem, the two choices for the Newton residual are $\boldsymbol{G}(\boldsymbol{v}_{p}^{n+1},p_{e,ijk}^{n+1})$ (black $+$) and $\boldsymbol{H}(\boldsymbol{E}_{ijk}^{n+1/2},p_{e,ijk}^{n+1})$. For the latter, the particles are advanced with the Picard-iterated implicit Boris push as described in Section~\ref{subcycleorbitav}, but, for a fair comparison between the two methods, we fix the number of particle sub-steps $N_{\nu p} = 1$. For this implementation, the results are plotted for a particle orbit convergence tolerance of $\epsilon_p = 10^{-4}$ (red $\times$), $\epsilon_p = 10^{-8}$ (green $\cdot$), and $\epsilon_p = 10^{-12}$ (blue $\odot$). The left panel shows how the momentum error, $\textrm{max}|(\boldsymbol{P}^n-\boldsymbol{P}^0)\cdot \boldsymbol{\hat{x}}|$, reduces with the L$^2$-norm of the residual in each implementation. For the first implementation with residual $\boldsymbol{G}(\boldsymbol{v}_{p}^{n+1},p_{e,ijk}^{n+1})$, the momentum is conserved to numerical round-off error regardless of the size of the residual norm. However, as discussed above, such an implementation is impractical for multi-scale problems. For the second implementation with residual $\boldsymbol{H}(\boldsymbol{E}_{ijk}^{n+1/2},p_{e,ijk}^{n+1})$ the momentum error depends on both the size of the residual and the particle convergence tolerance. For a loose particle tolerance of $\epsilon_p = 10^{-4}$, errors from the particle advance become dominant for $||residual||<10^{-10}$, where the momentum error becomes independent of residual norm. For tighter particle convergence tolerances of $10^{-8}$ and $10^{-12}$, the particle error is negligible and the momentum error reduces with the residual norm down to the level of numerical round-off. The relative total energy error, $\textrm{max}[(K_i + K_e)^n - (K_i + K_e)^0]/(K_i + K_e)^0$, is shown in the right panel for the same simulations. In contrast to the momentum error, there is strong dependence of the energy error on the non-linear convergence for both implementations. For the loose particle tolerance $\epsilon_p = 10^{-4}$, errors from the particles can dominate the energy error for $||residual|| < 10^{-6}$, but for the tighter tolerances the energy error can be converged down to numerical round-off. Note that small differences in the residual norm between implementations are due to different sized discrete jumps with each non-linear iteration, and should not be taken to be significant. We fix the particle convergence tolerance to $\epsilon_p = 10^{-13}$ to ensure that the errors are controlled by the outer Newton iteration. This typically requires about 4 Picard iterations per particle sub-step at every function evaluation.

\section{Explicit non-energy-conserving schemes for cold beam test}

\subsection{Momentum-conserving leapfrog scheme}\label{explicitmomentum}

For 1D-1V electrostatic problems, it is straightforward to implement an explicit momentum-conserving scheme directly using a leapfrog method for the particle advance, and a central-difference discretization of the electric field~\cite{rambo95}. This model is used for comparison in Section~\ref{coldbeam}, to demonstrate the favorable stability properties of the implicit conservative scheme. The known particle positions and velocities are staggered in time as $x_p^n$ and $v_{xp}^{n-1/2}$. First, the density moment is collected at cell centers as
\begin{equation}\label{nexplicit}n_i^n = \sum_p S(x_i - x_p^n)/\Delta x.\end{equation}
The electric field is then evaluated statically using central differences as
\begin{equation}\label{Eexplicit}E_i^n = -\frac{p_{ei+1}^n - p_{ei-1}^n}{en_i^n 2\Delta x},\end{equation}
with electron pressure $p_{ei}^n = T_{e0} (n_i^n)^\gamma$ for constant electron temperature $T_{e0}$ and ratio of specific heats $\gamma$. The particle velocities and positions are finally updated as
\begin{equation}\frac{v_{px}^{n+1/2} - v_{px}^{n-1/2}}{\Delta t} = \frac{q_p}{m_p} \sum_iE_i^n S(x_i-x_p^n),\end{equation}
\begin{equation}\frac{x_p^{n+1} - x_p^n}{\Delta t} = v_{px}^{n+1/2}.\end{equation}

It can be shown that a commonly used hybrid scheme~\cite{winske03,karimabadi04}, which uses forward-in-time extrapolation of the ion velocity moment to calculate the time advanced electric field, reduces to this scheme in the limit of 1D-1V. However, the momentum conserving property does not hold in multiple dimensions with a non-zero magnetic field.


\subsection{Explicit predictor-corrector scheme}\label{explicitpredcorr}

Several explicit predictor-corrector type schemes have been constructed for the hybrid model~\cite{harned82, winske86, kunz14}, which partly differ in how the time advanced fields are obtained for the prediction step. Here, we choose a scheme similar to Ref.~\cite{kunz14}, which avoids direct forward extrapolation of the fields by using an additional prediction-stage particle push. In the 1D-1V electrostatic limit, the known quantities at $t=n\Delta t$ are the particle position and velocity $x_p^n$ and $v_p^n$ respectively. The particle push for the prediction step uses the electric field calculated as in Eq.~(\ref{Eexplicit}), as
\begin{equation}x_p^* = x_p^n + \tfrac{\Delta t}{2}v_p^n,\end{equation}
\begin{equation}v_{p}^{n+1,\textrm{pred}} = v_p^n + \frac{q_p\Delta t}{m_p}\sum_i E_i^n S(x_i-x_p^*),\end{equation}
\begin{equation}x_{p}^{n+1,\textrm{pred}} = x_p^* + \tfrac{\Delta t}{2} v_{p}^{n+1,\textrm{pred}}.\end{equation}
The density moment is then gathered as
\begin{equation}n_i^{n+1,\textrm{pred}} = \sum_p S(x_i - x_p^{n+1,\textrm{pred}})/\Delta x,\end{equation}
and used to evaluate the predicted electric field as
\begin{equation}E_i^{n+1,\textrm{pred}} = -\frac{p_{ei+1}^{n+1,\textrm{pred}} - p_{ei-1}^{n+1,\textrm{pred}}}{en_i^{n+1,\textrm{pred}} 2\Delta x},\end{equation}
where $p_{ei}^{n+1,\textrm{pred}} = T_{e0} (n_i^{n+1,\textrm{pred}})^\gamma$. 

A correction step particle push is then performed using the time averaged electric field 
\begin{equation}\label{Eaveraged}E_i^{n+1/2,\textrm{pred}} = \tfrac{1}{2}(E_i^{n+1,\textrm{pred}} + E_i^n),\end{equation}
as
\begin{equation}v_{p}^{n+1} = v_p^n + \frac{q_p \Delta t}{m_p} \sum_iE_i^{n+1/2,\textrm{pred}} S(x_i-x_p^*),\end{equation}
\begin{equation}x_{p}^{n+1} = x_p^* + \tfrac{\Delta t}{2} v_{p}^{n+1},\end{equation}
to give the corrected particle position and velocity at the new time-step. This method does not conserve momentum exactly due to the form of the electric field in Eq.~(\ref{Eaveraged}), and because the gather and scatter operations are performed at different particle positions.



\bibliographystyle{elsarticle-num}
\bibliography{hybridcart.bib}







\end{document}